\documentclass[%
 reprint,
 superscriptaddress,
 amsmath,amssymb,
 aps,
 prl,
]{revtex4-1}

\usepackage[usenames,dvipsnames]{pstricks}
\usepackage{epsfig}
\usepackage{subfigure}
\usepackage{enumerate}

\usepackage{graphicx}
\usepackage{bm}
\usepackage{dcolumn}
\usepackage{color,soul}
\usepackage{xcolor}

\usepackage[utf8]{inputenc}
\usepackage[english]{babel}

\newcommand{\A}[1]{\Hat{#1}}

\newcommand{\T}[1]{\widetilde{#1}}
\newcommand{\pd}[2]{\dfrac{\partial #1}{\partial #2}}

\newcommand{\vp}[1]{\mathbf{#1}}

\newcommand{\abs}[1]{\left| #1\right|}
\newcommand{\ket}[1]{|#1\rangle}
\newcommand{\bra}[1]{\langle #1| }

\newcommand{\Ex}[1]{\langle #1\rangle}

\newcommand{\OO}[1]{\overline{\overline{#1}}}

\begin{document} 

\title{Anomalous optical saturation of low-energy Dirac states in graphene and its implication for nonlinear optics}

\author{Behrooz Semnani}
\email{bsemnani@uwaterloo.ca}
\affiliation{Department of Physics, Chalmers University of Technology, G\"oteborg, Sweden}
\affiliation{Institute for Quantum Computing (IQC), Waterloo, ON, Canada}
\affiliation{Waterloo Institute for Nanotechnology, University of Waterloo, Waterloo, ON, Canada}
\affiliation{Department of Electrical \& Computer Engineering, University of Waterloo, Waterloo, ON, Canada}

\author{Roland Jago}
\affiliation{Department of Physics, Chalmers University of Technology, G\"oteborg, Sweden}

\author{Safieddin Safavi-Naeini}
\affiliation{Waterloo Institute for Nanotechnology, University of Waterloo, Waterloo, ON, Canada}
\affiliation{Department of Electrical \& Computer Engineering, University of Waterloo, Waterloo, ON, Canada}

\author{Amir Hamed Majedi}
\affiliation{Waterloo Institute for Nanotechnology, University of Waterloo, Waterloo, ON, Canada}
\affiliation{Department of Electrical \& Computer Engineering, University of Waterloo, Waterloo, ON, Canada}
\affiliation{Perimeter Institute for Theoretical Physics (PI), Waterloo, ON, Canada}

\author{Ermin Malic}
\affiliation{Department of Physics, Chalmers University of Technology, G\"oteborg, Sweden}
 
\author{Philippe Tassin}
\affiliation{Department of Physics, Chalmers University of Technology, G\"oteborg, Sweden}

\begin{abstract}
We reveal that optical saturation of the low-energy states takes place in graphene for arbitrarily weak electromagnetic fields. This effect originates from the diverging field-induced interband coupling at the Dirac point. Using semiconductor Bloch equations to model the electronic dynamics of graphene, we argue that the charge carriers undergo ultrafast Rabi oscillations leading to the anomalous saturation effect. The theory is complemented by a many-body study of the carrier relaxations dynamics in graphene. It will be demonstrated  that the carrier relaxation dynamics is slow around the Dirac point, which in turn leads to a more pronounced saturation. The implications of this effect for the nonlinear optics of graphene are then discussed. Our analysis shows that the conventional perturbative treatment of the nonlinear optics, i.e., expanding the polarization field in a Taylor series of the electric field, is problematic for graphene, in particular at small Fermi levels and large field amplitudes.
\end{abstract}

\maketitle

Graphene is a two-dimensional material made of carbon atoms in a honeycomb structure. Its reduced dimensionality and the symmetries of its crystalline structure render graphene a gapless semiconductor~\cite{RevModPhys.81.109}. Graphene exhibits a wealth of exceptional properties, including a remarkably high mobility at room temperature~\cite{BOLOTIN2008351}, Klein tunneling and Zitterbewegung~\cite{Zitterbewegung,Young2009}, existence of a nonzero Berry phase, anomalous quantum Hall effect~\cite{Zhang2005,PhysRevLett.95.146801,Gorbachev1254966}, quantum-limited intrinsic conductivity~\cite{Novoselov2005}, and a unique Landau level structure~\cite{PhysRevLett.100.206801,Mittendorff2014}. Underlying these peculiar electronic properties are its pseudo-relativistic quasiparticles that obey the massless Dirac equation~\cite{RevModPhys.81.109}. As a direct consequence of their massless nature, the Dirac fermions have definite chiralities~\cite{Katsnelson2006,pal2011dirac}. Owing to the specific symmetries of the crystalline structure of graphene, the dynamics of the massless Dirac quasiparticles and their chiral character are topologically preserved---i.e., many-body induced band renormalizations as well as any moderate perturbations of the lattice will not open a gap in graphene's band structure~\cite{bernevig2013topological}. A large number of the  unusual properties of graphene are associated with the topologically protected band-crossing and the chiral dynamics of the charge carriers~\cite{Zitterbewegung}.

One major consequence of the topologically protected chirality of the charge carriers is the anomalous structure of the interband coupling mediated by an electromagnetic field. Its dipole matrix element obtained in the length gauge~\cite{Aversa1995} exhibits a singularity at the degeneracy points, in contrast to ordinary (and even other gapless) semiconductors~\cite{Behrooz,Kelardeh2014}. This has raised some controversy regarding the treatment of the optical response of graphene~\cite{Behrooz,Cheng2015,Hipolito2018}. Specifically, the perturbative treatment of the nonlinear optical response has been questioned~\cite{Cheng2015,Cheng2014}. The nonlinear optical coefficients of graphene obtained by means of perturbation theory suffer from a nonresolvable singularity~\cite{Cheng2015,Behrooz}. Although substantial effort has been spent on developing comprehensive models for the nonlinear optical response of graphene \cite{Cheng2014,Cheng2015,Cheng2015_Numeric,Christensen2015,Mikhailov2016,Mikhailov2017,Winzer2017}, a self-consistent theoretical model that can resolve the above issue is still lacking. In addition, many experimental studies of the nonlinear optics of graphene have been reported~\cite{Zhang2012,Chen2013,Hartschuh2015,Dremetsika2016,Vermeulen2016,Behrooz_experimental}---some of these studies are difficult to reconcile with existing theoretical models.

In this Letter, we show that the singular nature of the interband dipole coupling has some significant physical implications: it causes the charge carriers in the vicinity of the Dirac points to undergo ultrafast Rabi oscillations accompanied by slow relaxation dynamics, which, intriguingly, yields an anomalous saturation effect. This finding necessitates revisiting the perturbative treatment of the nonlinear optical response of graphene to account for the extreme nonlinear interactions around the Dirac points. These conclusions will be reached by describing the dynamics of the charge carriers with semiconductor Bloch equations (SBEs)~\cite{Lindberg1988,Avetissian2013}.

We consider a free-standing graphene monolayer (in the $xy$ plane) illuminated by a normally incident electromagnetic field. The monochromatic and spatially uniform optical field at the graphene layer is described by $\vp{E}(t)=\vp{E}_0e^{i\omega t}+\mathrm{c.c.}$, where $\vp{E}_0$ is parallel to the graphene plane. 
The light-matter interaction is considered semiclassically and the external field coupling is obtained in the length gauge~\cite{Aversa1995}. For photon energies below approximately $2\mathrm{eV}$, the electronic dynamics of the  quasiparticles in the absence of external radiation is adequately described by the massless Dirac equation yielding the relativistic energy-momentum dispersion $\mathcal{E}_\vp{k}=\pm\hbar v_F \abs{\vp{k}}$~\cite{RevModPhys.81.109}, where $\vp{k}$ is the Bloch wave vector with respect to the Dirac point and $v_F$ is the Fermi velocity. 

The SBEs describe the coupled dynamics of the population difference $\mathcal{N}(\vp{k},t)$ and the polarization (coherence) $\mathcal{P}(\vp{k},t)$ in the momentum state \textbf{k}. In the absence of electromagnetic radiation, the population difference relaxes to $\mathcal{N}^{eq}_\vp{k}=f(\hbar v_F k)-f(-\hbar v_F k)$, where $f(\mathcal{E})$ is the Fermi-Dirac distribution. An electromagnetic field drives the system out of equilibrium via the coupled intraband and interband dynamics. In a moving frame $\{\tau,\vp{k}'\}=\{t,\vp{k}-\delta\vp{k}(t)\}$, where $\delta\vp{k}$ obeys $\frac{\partial\delta\vp{k}}{\partial t}+\Gamma\delta\vp{k}=-\frac{e}{\hbar}\vp{E}(t)$  ($\Gamma$ is a phenomenological intraband relaxation coefficient),
the dynamics of the charge carriers is governed by
\begin{subequations}
\begin{multline}\label{eq:SBE1}
\pd{\mathcal{N}({\vp{k}'},\tau)}{\tau}=-\gamma^{(1)}_{\vp{k'}}(\mathcal{N}({\vp{k}'},\tau)-\mathcal{N}^{eq}_{\vp{k}'}) \\
 -2\Phi(\vp{k}',\tau)\mathrm{Im}\left\{\mathcal{P}(\vp{k}',\tau)\right\},
\end{multline}
\begin{multline}\label{eq:SBE2}
\pd{\mathcal{P}({\vp{k}',\tau)}}{\tau} =-\gamma^{(2)}_{\vp{k}'}\mathcal{P}(\vp{k}',\tau)+\\
 i\varpi_{\vp{k}'}\mathcal{P}(\vp{k}',\tau)+\frac{i}{2} \Phi(\vp{k}',\tau)\mathcal{N}(\vp{k}',\tau),
\end{multline}
\end{subequations}
where $\Phi(\vp{k},t)=\frac{e\vp{E}\cdot\A{\varphi}_{\vp{k}}}{\hbar k}$ is the matrix element of the external potential of the direct optical transition, and the unit vector $\A{\varphi}_{\vp{k}}$ is defined as $\A{\varphi}_{\vp{k}}=\A{z}\times\vp{k}/k$. The frequency $\hbar\varpi_{\vp{k}}=2\mathcal{E}_{\vp{k}}$ is the energy difference between the energy levels of the conduction and valence bands. $\gamma^{(1)}_{\vp{k}}$ and $\gamma^{(2)}_{\vp{k}}$ are $k$-dependent relaxation coefficients stemming from many-body effects such as electron-electron and electron-phonon interactions. It is worth pointing out that the moving frame accounts for the coherent Bloch oscillations, which play an important role in high-harmonics generation~\cite{Yoshikawa}. The detailed derivation of our theoretical model is provided in the Supplemental Material~\cite{SupplMat}. 

The light-graphene interaction as described by Eqs.~(\ref{eq:SBE1})-(\ref{eq:SBE2}) can be interpreted as an ensemble of inhomogeneously broadened two-level systems (one for each $\mathbf{k}$). The last term in each of the two equations will lead to Rabi oscillations. Because of the singularity in $\Phi(\vp{k}',\tau)$ for $|\mathbf{k}|\to0$, we can expect ultrafast Rabi oscillations around the Dirac point, which are damped by many-body interactions. The decay terms drive the two-level systems towards an equilibrium state. Since the interband coupling is strong around the Dirac point (equivalent to highly intense illumination), the effective field leaves the two-level systems in a statistical mixture of the ground and excited states with equal weights and absorption quenching takes place. Thus, the states around the Dirac points undergo a  saturation effect, even when illuminated by an arbitrarily weak electromagnetic field.

\begin{figure*}
\includegraphics[width=0.98\textwidth]{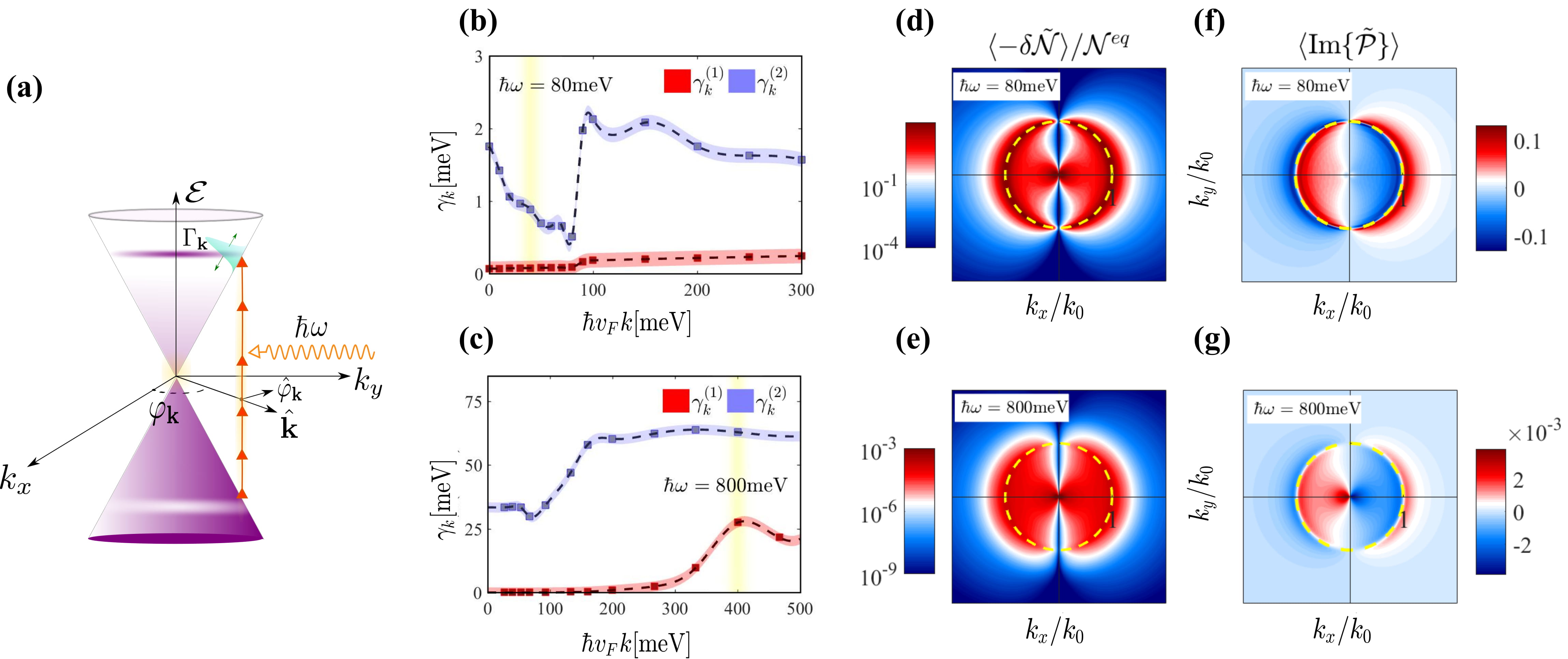}
\caption{\footnotesize  
(a) Low-energy band structure of graphene. The angle $\A{\varphi}_{\vp{k}}$ and the magnitude of the Bloch wavenumber $k$ are defined in polar coordinate system in reciprocal space.  (b)\& (c) k-dependent population and coherence relaxation coefficients shown by $\gamma_k^{(1)}$ and $\gamma_k^{(2)}$ respectively. The coefficients are calculated  for $\hbar\omega=80\mathrm meV$ and $800\mathrm{meV}$ electromagnetic excitation. In both cases the electric field magnitude is $10^6 \mathrm{V/m}$ and is polarized along the y-axis.  (d) \& (e) Relative change in the population difference with respect to the equilibrium $\delta\mathcal{N}=\mathcal{N}-\mathcal{N}_{eq}$ for $\hbar\omega=80\mathrm meV$ and $800\mathrm{meV}$, respectively. $k_0$ is defined via $\hbar v_F k_0=\hbar\omega/2$  (f)\& (g) The corresponding steady-state polarization.}
\label{fig:saturation_time}
\end{figure*}

This saturation behavior can be further understood by studying the steady-state solution of the SBEs, which is
\begin{equation}
\T{\mathcal{N}}_{\vp{k}}^{st}=\mathcal{N}^{eq}_{\vp{k}}\frac{\gamma^{(1)}_{\vp{k}}}{\gamma^{(1)}_{\vp{k}}+\gamma^{(2)}_\vp{k}\abs{\T{\Phi}_\vp{k}}^2/ \abs{\gamma^{(2)}_{\vp{k}}+i\Delta_{\vp{k}}}^2},
\label{eq:Nst}
\end{equation}
where $\T{\Phi}_{\vp{k}}={e\vp{E}_0\cdot\A{\varphi}_{\vp{k}}}/\hbar k$ is the complex phasor associated with ${\Phi(\vp{k},t)}$. The function $\Delta_{\vp{k}}=\omega-\varpi_\vp{k}$ denotes the detuning of the two-level system at $\vp{k}$ with respect to the excitation.

Since $|\T{\Phi}_{\vp{k}}|$ is arbitrarily large for small-k states, in the vicinity of the Dirac point,  the population $\T{\mathcal{N}}_{\vp{k}}$ cannot be expanded in a Taylor series of the field $\tilde{\Phi}$. This implies that the nonlinear optics of graphene is in principle a nonperturbative problem. Indeed, due to the singularity of the interband coupling in graphene, there is always a region around the Dirac point where graphene is optically saturated. The saturation threshold $E^{sat}_\vp{k}$ is given by
\begin{equation}\label{eq:Es}
eE^{sat}_\vp{k}=\hbar k\sqrt{\Delta^2_{\vp{k}}\frac{\gamma^{(1)}_\vp{k}}{\gamma^{(2)}_\vp{k}}+\gamma^{(1)}_\vp{k}\gamma^{(2)}_\vp{k}}.
\end{equation}
Saturation occurs, of course, in any two-level system at high field intensities. However, in graphene the saturation threshold field $E^{sat}_\vp{k}$ is zero at the Dirac point ($k\to 0$) and, hence, there is always a region of k-space where $E_0>E^{sat}_\vp{k}$, even for arbitrarily weak intensities.

The peculiar low-threshold saturation mechanism in graphene can be quantitatively resolved using a time-domain analysis of the graphene SBEs. For the sake of comparison, this analysis has been performed for two distinct continuous excitations with optical frequency of $\hbar\omega=80\mathrm{meV}$ (terahertz range) and $\hbar\omega=800\mathrm{meV}$ (infrared), respectively. In both cases, the electric field is linearly polarized along the $\A{y}$ direction with magnitude $E_0=10^6\mathrm{V/m}$. Graphene is assumed to be undoped here and is initially held at room temperature. The relaxation coefficients $\gamma^{(1)}_{\vp{k}}$ and $\gamma^{(2)}_{\vp{k}}$ are determined using a microscopic theory, which encompasses carrier-carrier as well as carrier-phonon scattering channels and takes into account all relevant relaxation paths including interband and intraband and even inter-valley processes~\cite{Malic2011,malic2013graphene}. The reader is referred to the Supplemental Material for the details of the many-body model~\cite{SupplMat} as well as the methodology used to extract the relaxation coefficients. The resulting relaxation coefficients are plotted in Fig.~\ref{fig:saturation_time}(b) and (c). We note in particular that $\gamma^{(1)}_{\vp{k}}$ tends to be zero around the Dirac point, which confirms the slow relaxation dynamics suggested above.

The relative change in the stationary component of the population difference due to the optical excitation as well as the amplitude of the oscillating induced polarization are shown in Fig.~\ref{fig:saturation_time}(d) and (e). To obtain the steady-state components, we performed Fourier analysis within a time window where the transient response has died out. As expected, a well-pronounced modified population difference around the Dirac point due to the spontaneous polarization effect (dark red region around the center) is observed. This effect is stronger for lower-frequency excitations---indeed, according to Eq.~\eqref{eq:Es} a smaller detuning yields a weaker saturation threshold. The region in k-space where the spontaneous optical saturation is significant is well extended from the Dirac point. We note here that the size of the region depends on the applied field intensity---we will show below that this is the origin of the nonperturbative nature of the nonlinear optical response. In addition, there is, of course, the traditional optical saturation region for $\Delta_{\vp{k}}\approx 0$ (indicated by the yellow dashed line).

Before we continue with the importance of this anomalous optical saturation for the nonlinear optics of graphene, let us briefly make a few remarks. First, the origin of the inverse dependence of the interband transition matrix element on the wavenumber can be linked to the distinctive mathematical structure of the current operator. It is straightforward to show that the interband coupling matrix element at wavenumber $\vp{k}$ is  $\A{\vp{r}}_{cv}\approx\frac{i\hbar}{ev_F}\vec{\mathcal{J}}_{cv}(\vp{k})/[\mathcal{E}_c(\vp{k})-\mathcal{E}_v(\vp{k})]$ where $\vec{\mathcal{J}}_{cv}$ is the off-diagonal element of the current operator. In contrast to ordinary semiconductors, the off-diagonal components of the current operator in graphene and other chiral materials are strictly nonzero even at the band crossing points \cite{Auslender2007}. As a direct consequence of this property of massless Dirac quasiparticles, the interband part of the position operator carries a first-order singularity at the degeneracy point. Second, one may wonder why we have not used the velocity gauge, in which optical coupling is obtained by minimal substitution $\hbar\vp{k}\to\hbar\vp{k}+e\vp{A}$ where $\vp{E}=-\partial{\vp{A}}/\partial{t}$. It is important to note that this approach is not gauge invariant in the ``effective Hamiltonian'' picture \cite{Virk2007,Pedersen2017,Ventura2017}. We show in the Supplemental Material that a modification of the velocity gauge is indeed required to yield a physically correct result~\cite{SupplMat}. This modification gives rise to the $1/k$ dependence of the interband coupling in the vicinity of the Dirac point~\cite{Ventura2017}. Third, although SBEs can model the essential physics, a number of approximations have evidently been made in deriving them. For instance, for intense sub-terahertz excitations, quasi-instantaneous thermal effects can cause a population pulsation at a rate faster than the interband Rabi oscillations and thus they cannot be modelled by a spectral broadening~\cite{Hafez}. However, owing to the fact that the anomalous saturation effect occurs around the Dirac point where the quasiparticles are off-tuned with the excitation photons, quasi-instantaneous thermal effects---taking place dominantly around the zero-detuning circle---have  minimal impact on the anomalous saturation effect.  Moreover, since around the Dirac point population difference is fairly independent of temperate $T$, i.e.,  $\partial{\mathcal{N}^{eq}_{\mathbf{k}}}/\partial{T}\approx 0$, temporal fluctuations of the temperature will not significantly obscure the region where saturation happens.

\begin{figure}[h]
\includegraphics[width=0.49\textwidth]{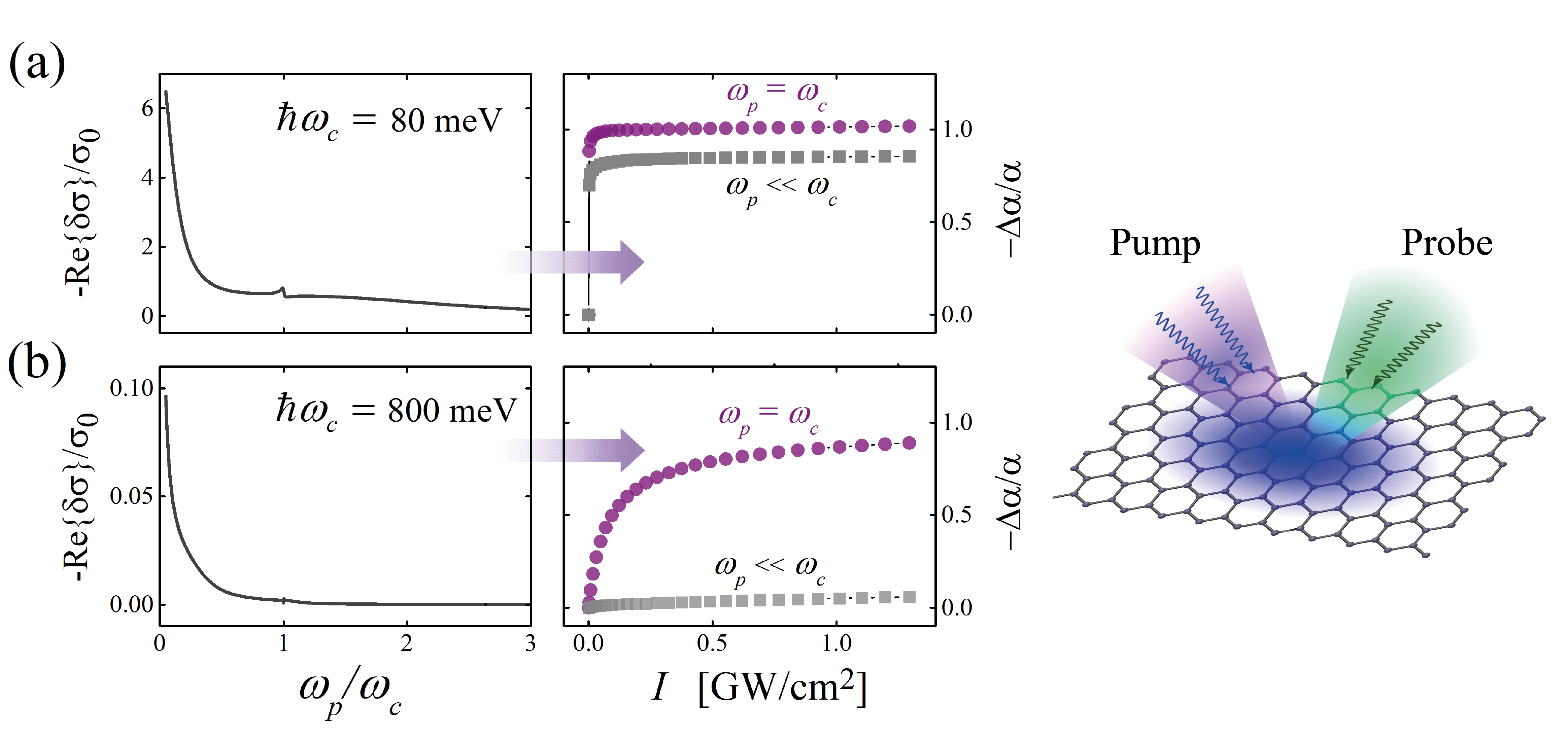}
\caption{ \footnotesize Change in the conductivity of graphene seen by the probe field (with frequency $\omega_p$) in a pump-probe experiment for the pump frequencies (a) $\hbar\omega_c=80\mathrm{meV}$ and (b) $\hbar\omega_c=800\mathrm{meV}$. The conductivity is normalized to $\sigma_0=e^2/4\hbar$.   Corresponding changes of the absorption coefficient of graphene for different intensities of the pump field are shown to the right of the conductivity plots.}
\label{fig:pump-probe}
\end{figure}

\begin{figure*}[t]
\centering
\includegraphics[width=0.92\linewidth]{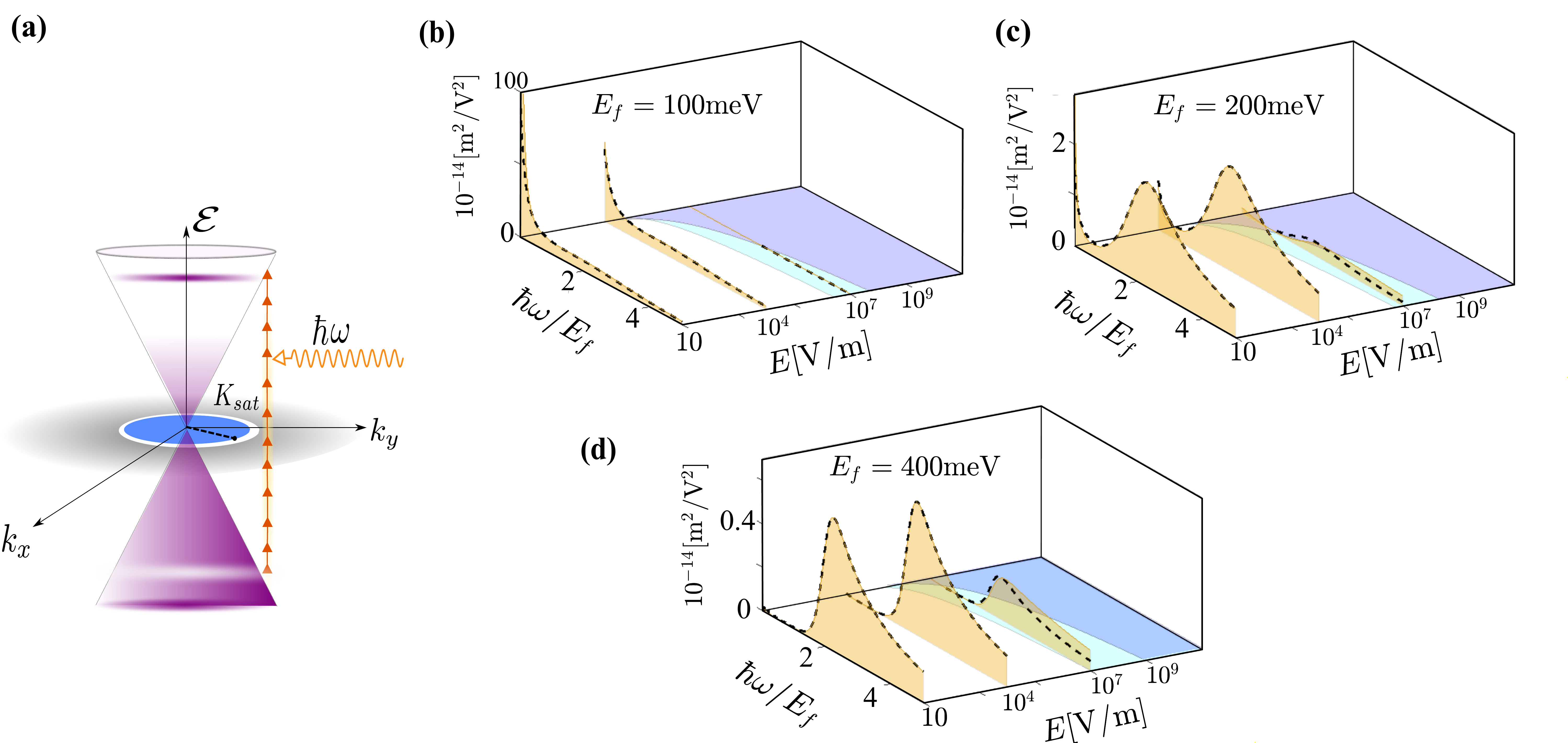}
\caption{\footnotesize (a)~In the calculation of the semiperturbative nonlinear coefficient, the saturated region displayed by the dark blue disc is excluded (b-d) Kerr-type nonlinearity of graphene obtained from the analytical nonperturbative approach (yellow shaded curves) and the semiperturbative approach (i.e., $|\sigma^{(3)}(\omega,\omega,-\omega)|$, black dotted curves) plotted for different field magnitudes (i.e. $E$). The z-axis corresponds to  $|\sigma^{(3)}(\omega,\omega,-\omega)|/\sigma_0$ where $\sigma_0=e^2/4\hbar$.   The two distinctive blue shaded regions are the saturation regions in $E-\omega$ plane due to, respectively, zero-detuning (light blue) and the strong interband coupling in the vicinity of the Dirac points (dark blue). Results for different Fermi levels ($E_f=100,200,400\mathrm{meV}$) are displayed in (b-d).
For low Fermi levels the nonlinear optical response is noticeably field-dependent. if the photon energy lies on the dark-blue disk semiperturbative coefficients would not exist and  thus the black dotted curves are not extended within the low-energy saturation domain. In the zero-detuning saturation domain (light blue area), the semiperturbative analysis fails and it cannot follow the non-perturbative results.}
\label{fig:semi-perturbative}
\end{figure*}

We now turn to the nonlinear optics of graphene. Let us consider a nonlinear pump-probe experiment in which graphene is simultaneously subjected to the a pump ($\omega_c$) and a weaker probe ($\omega_p$) laser beam. The conductivity tensor of graphene in the presence of the pump field and ``seen'' by the probe field is calculated in the Supplementary Material~\cite{SupplMat}. Fig.~\ref{fig:pump-probe} displays the change to the conductivity tensor due to the pump fields described earlier. The real part of the conductivity is related to the absorption coefficient of the probe beam via $\alpha \approx \mathrm{Re}\{\sigma\}/4\varepsilon_0c$, where $c$ is the speed of light in vacuum. The relative change in absorption of the probe beam is also shown in Fig.~\ref{fig:pump-probe}. For the $80 \mathrm{meV}$ pump beam, there is strong saturation for a probe beam at the same frequency---this is because the pump beam has saturated the interband transition. For a probe beam with a much lower frequency ($\omega_p \approx \omega_c/100$), there is also strong saturation---which must be due to the unconventional effect discussed above. For the $800 \mathrm{meV}$ pump, a weaker saturation is observed for a low-frequency probe. Indeed, for a higher-frequency pump the region of the low-threshold saturation is smaller [observe in Eq.~(\ref{eq:Es}) that a larger detuning results in a larger saturation field]. Although the above example discusses the anomalous saturation effect for undoped graphene, the effect happens for a range of Fermi levels. Therefore, it will also be observed in samples where the charge-neutrality point fluctuates in space.

We now move on to discuss the applicability of perturbation theory in the analysis of the nonlinear response of graphene. As detailed in Ref.~\cite{Behrooz}, the standard perturbative treatment of the optical response of graphene leads to a nonresolvable singularity in its higher-order nonlinear coefficients (beyond the linear response) originating from the small-$\mathbf{k}$ states. We now understand that these states should be excluded because they are saturated. However, since the saturation threshold depends on the field intensity, the nonlinear response coefficients calculated with standard perturbation theory become field dependent too. But let us investigate under what circumstances the standard nonlinear coefficients have meaning.

As mentioned earlier, neglecting the momentum of the absorbed photon, the optical transitions are vertical in k-space and, therefore, every point in the reciprocal space can be treated independently. It is argued in Refs.~\cite{Behrooz,Cheng2015,Mikhailov2016} that nonlinear frequency mixing in graphene can be decomposed into a number of additive contributions (see the Supplemental Material for a complete theoretical analysis~\cite{SupplMat}), i.e., the nonlinear conductivity tensor is
$\smash{\overline{\overline{\sigma}}}\vphantom{\sigma}^{(l)}\approx\sum_{\mathbf{k},\abs{\mathbf{k}}>K_\mathrm{sat}}\smash{\overline{\overline{\mathcal{I}}}}\vphantom{\mathcal{I}}_{\mathbf{k}}^{(l)}$, 
where $K_{sat}$ is the radius (with respect to the Dirac point) of the spontaneously saturated region in k-space and is obtained from Eq.~\eqref{eq:Es}:
\begin{equation}\label{eq:Ksat}
\hbar v_F K_\mathrm{sat}\approx\frac{1}{2}\hbar\omega-\sqrt{(\frac{1}{2}\hbar\omega)^2-2\hbar v_FeE_0\sqrt{\frac{\gamma^{(2)}}{\gamma^{(1)}}}},
\end{equation}
and $\smash{\overline{\overline{\mathcal{I}}}}\vphantom{\mathcal{I}}^{(l)}_{\vp{k}}$  represents the contribution of the quasiparticles with Bloch index $\vp{k}$  to the $l$'th order nonlinear optical response. $E_0$ is the magnitude of the largest electric field component (most often a pump field) participating in the nonlinear process and $\omega$ is its frequency. 

In order to gain insight into the intensity dependence of the nonlinear response coefficients obtained from the above-described semiperturbative approach, we compare in Fig.~\ref{fig:semi-perturbative} the third-order nonlinear response defined as $|\sigma_{xxxx}^{(3)}(\omega,\omega,-\omega)|$ (where we have now used k-independent relaxation constants as usual) with the results of the full solution of SBEs (see the Supplemental  Material~\cite{SupplMat}). The yellow shaded curves display the nonperturbative solution and the black dotted curves are the results of semiperturbative approach. First, owing to the low-threshold  saturation effect, there is a noticeable field dependence of the third-order nonlinear (Kerr-like) response for lower Fermi energies. When the field intensity becomes large enough to extend the saturation region to the excited k-states, the semiperturbative approach fails. As the Fermi energy becomes larger, the optically induced Pauli blocking becomes less important as the low-energy states are already Pauli blocked. It is worth pointing out that we have observed significant dependence of the results on $K_{sat}$. Therefore, the exact exclusion of the saturated region is necessary to achieve accurate results. 

In conclusion, we have demonstrated that the topologically protected singular interband coupling in graphene leads to ultrafast Rabi oscillations, exciting the quasiparticles faster than they can relax back to the ground state. This leads to an anomalous optical saturation of the low-energy quasiparticles in graphene. Subsequently, we have shown that due to this effect the small-k states have to be excluded for the perturbative calculation of the nonlinear optical coefficients of graphene. As a result, the nonlinear coefficients obtained from perturbation theory exhibit noticeable field dependence, particularly for small Fermi levels. {Although the effect revealed in this Letter has not yet been observed directly, several experiments have demonstrated the nonperturbative nature of the nonlinear optical response in graphene.} For instance, Refs.~\cite{Zhang2012} and \cite{Behrooz_experimental}, which {present experimental results for the Kerr nonlinear coefficient of graphene obtained from z-scan measurements, demonstrate that the effective Kerr coefficient is not independent of the intensity of light. At very intense illuminations, the field dependence of the Kerr coefficient might of course have multiple origins, but this effect has been observed even in the weak-field regime}, e.g., in Ref.~\cite{Yoshikawa,Bowlan}, {which report recent experimental observations of optical and terahertz high-harmonic generation in doped and nearly-undoped graphene.} Our theory provides an explanation for these experimental results indicating the nonperturbative nature of the nonlinear optics of graphene. We speculate that similar effects may be found in other Dirac materials and in Weyl semimetals.

\begin{acknowledgments}
Work at Waterloo has been supported by the Natural Science and Engineering Research Council of Canada (NSERC) and the Canada First Research Excellence Fund. Work at Chalmers has been supported by the Rune Bernhardsson Graphene fund, by Vetenskapsr\aa{}det under grant No.\ 2016-03603, and by the European Union's Horizon 2020 research and innovation programme under grant agreement No.\ 785219.
\end{acknowledgments}

\newpage
\onecolumngrid

\section{Supplemental Material}

 \section{Derivation of Semiconductor Bloch Equations  for Graphene}

\subsection{Equations of Motion}
\begin{figure}[t]
\includegraphics[scale=0.06]{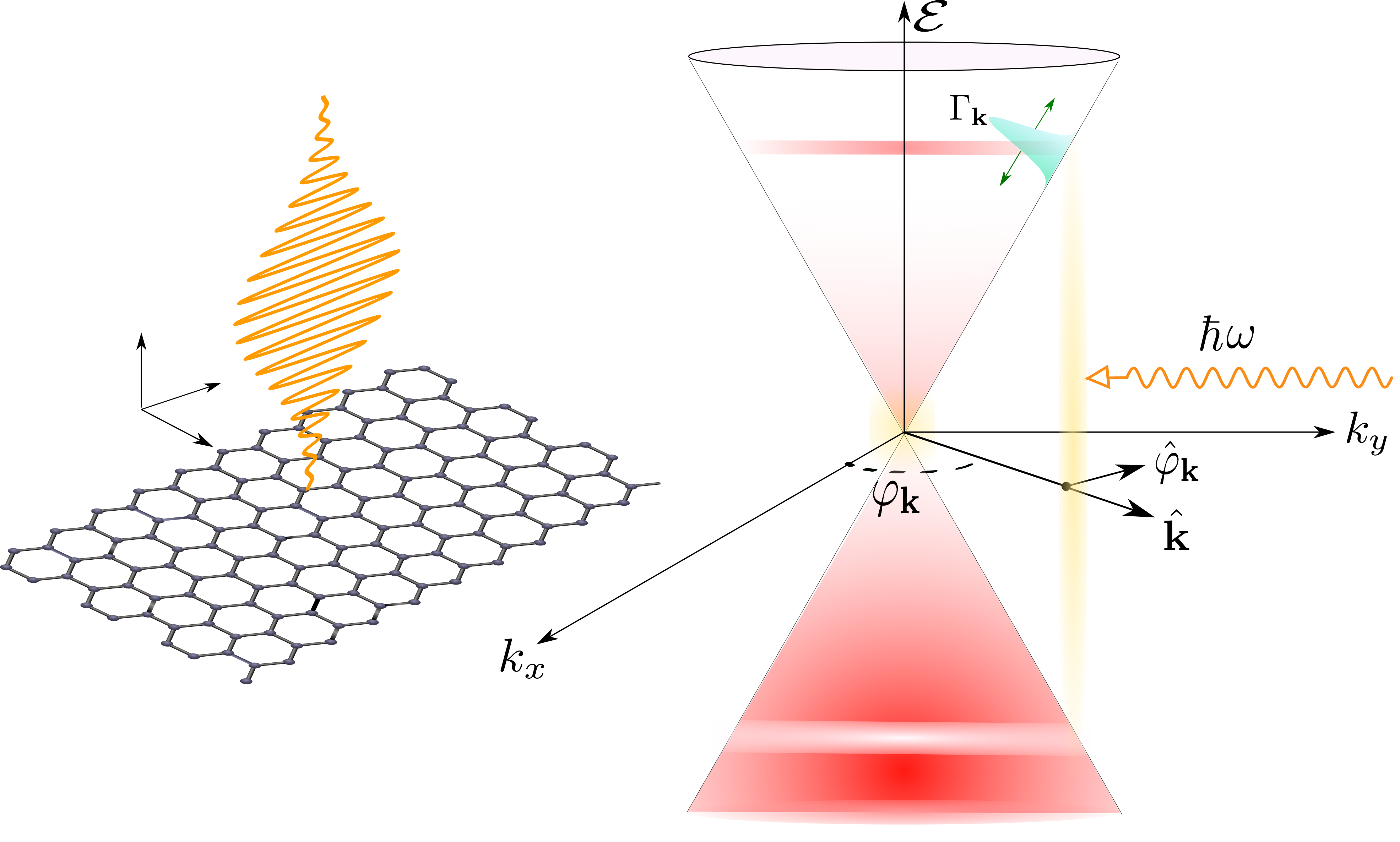}
\caption{Schematic of the effective-carrier dispersion in graphene. The vector $\vp{k}$ is the Bloch wavenumber with respect to the Dirac point. }
\label{fig:graphene_schematics}
\end{figure}

To begin,  we use the low-energy effective Hamiltonian of graphene, which is described by $\mathcal{H}_0=\hbar v_F \vp{k}\cdot\vec{\A{\sigma}}$ in the sublattice (pseudospin) basis.  Pauli matrices $\vec{\A{\sigma}}=\A{\sigma}_i\A{x}_i$ are  used to expand the operators. For the time being, we ignore the many-body effects such as Coulomb interactions and their collective influence will be phenomenologically included. The calculations are carried out within the  length gauge,  using the additive potential $V_{ex}=e\vp{E}\cdot\vp{r}$ to couple the electromagnetic field to the dynamical equations. The time evolution of the density matrix is then obtained through the Liouville equation as \cite{Behrooz}
\begin{equation}\label{eq:2000}
i\hbar\pd{\A{\rho}_{\vp{k}}}{t}=\hbar v_F \vp{k}\cdot [\A{\vec{\sigma}},\A{\rho}_{\vp{k}}]+ie\vp{E}\cdot \nabla_{\vp{k}}\A{\rho}_{\vp{k}}
\end{equation}
The $2\times 2$ pseudospin density matrix $\A{\rho}_{\vp{k}}$ can be expanded in terms of Pauli matrices
\begin{equation}\label{eq:101}
\A{\rho}_{\vp{k}}=n_{\vp{k}}\A{I}+\vec{m}_{\vp{k}}\cdot\A{\vec{\sigma}}
\end{equation}
Substituting Eq.~\eqref{eq:101} into Eq.~\eqref{eq:2000}, one obtains decoupled equations for coefficient $n_{\vp{k}}$ and vector $\vec{m}_{\vp{k}}$ (see Ref.~\cite{katsnelson}):
\begin{align}
&\pd{n_{\vp{k}}}{t}=\frac{e}{\hbar}\vp{E}\cdot \nabla_{\vp{k}} n_{\vp{k}}\label{eq:110}\\
&\pd{\vec{m}_{\vp{k}}}{t}=2v_F\left(\vp{k}\times \vec{m}_{\vp{k}}\right)+\frac{e}{\hbar}\vp{E}\cdot \nabla_{\vp{k}}\vec{m}_{\vp{k}}\label{eq:111}
\end{align} 
We will also need the current operator to construct the optical response functions
\begin{equation}
\A{\vec{\mathcal {J}}}_{\vp{k}} =-\frac{e}{\hbar}\pd{\A{\mathcal{H}}_{\vp{k}}}{\vp{k}}=-ev_F\A{\vec{\sigma}}
\end{equation}
and the current density becomes
\begin{equation}
\vp{J}=\Ex{\A{\vec{\mathcal {J}}}_{\vp{k}}}=\mathrm{Tr}\left(\A{\vec{\mathcal {J}}}_{\vp{k}}\A{\rho}_{\vp{k}}\right)=-2ev_F\left(\A{x}\A{x}+\A{y}\A{y}\right)\cdot\sum_{\vp{k}}\vec{m}_{\vp{k}}
\end{equation}
Having derived the equations of motion in the sublattice basis, we can now switch to the energy diagonal basis. We use ``$\sim$'' to denote the matrix representation of the operators in the valence and conduction basis. In the energy diagonal basis, matrices
$
\begin{pmatrix}
1 &0
\end{pmatrix}^T
$
and 
$ \begin{pmatrix}
0 &1 
\end{pmatrix}^T $
are adopted for the upper and the lower energy levels, respectively. In the energy diagonal basis, the density matrix and the current operator become
\begin{align}
\tilde{\rho}_{\vp{k}}=&\A{I}n_{\vp{k}}+\vec{m}_{\vp{k}}\cdot \left(\A{\vp{k}}\sigma_z +\A{\varphi}_{\vp{k}}\sigma_{y}-\A{\vp{z}}\sigma_x \right)\label{eq:rho100}\\
\tilde{\vec{\mathcal{J}}}_{\vp{k}}=&-ev_F\left(\A{\vp{k}}\sigma_z+\A{\varphi}_{\vp{k}}\sigma_y\right)\label{eq:J100}
\end{align}
where  $\A{\vp{k}}$ and $\A{\varphi}_{\vp{k}}$ are the unit vectors shown in Fig.~\ref{fig:graphene_schematics}. At thermal equilibrium, before switching on the incident field, the density distribution obeys Fermi-Dirac statistics
\begin{equation}
\Ex{\A{\xi}_{\vp{k}c}^\dagger\A{\xi}_{\vp{k}c}}_0 = f(\mathcal{E}(\vp{k})) \quad,\quad\Ex{\A{\xi}_{\vp{k}v}^\dagger\A{\xi}_{\vp{k}v}}_0 = f(-\mathcal{E}(\vp{k})) 
\end{equation} 
where $\A{\xi}_{\vp{k}v}$ and $\A{\xi}_{\vp{k}c}$ are the many-body annihilation operators for the valence and conduction bands, respectively. 
The subscript $0$ denotes the equilibrium state and $\mathcal{E}(\vp{k})=\hbar v_F k$ is the upper energy level. The distribution $f(E)$ is the Fermi-Dirac distribution function
\begin{equation*}
f(E)=\frac{1}{1+\exp\left(\frac{E-\mu}{k_BT}\right)}
\end{equation*}
where $T$ and $\mu$ are, respectively, the temperature and the chemical potential associated with the Fermi energy level $E_f$.  

\subsection{Semiconductor Bloch Equations}
It can be shown that the optical response under continuous excitation wave can be estimated based solely on the dynamics of (i) the microscopic population difference $\mathcal{N}(\vp{k},t)$ and (ii) the microscopic polarization $\mathcal{P}(\vp{k},t)$ defined as 
\begin{align}
&\mathcal{N}(\vp{k},t)=\Ex{\A{\xi}_{\vp{k}c}^\dagger\A{\xi}_{\vp{k}c}}-\Ex{\A{\xi}_{\vp{k}v}^\dagger\A{\xi}_{\vp{k}v}}=2\A{\vp{k}}\cdot\vec{m}\label{def:N}\\
&\mathcal{P}(\vp{k},t)=\Ex{\A{\xi}_{\vp{k}v}^\dagger\A{\xi}_{\vp{k}c}}=-\A{\vp{z}}\cdot\vec{m}+i\A{\varphi}_{\vp{k}}\cdot\vec{m}\label{def:P}
\end{align}
Taking $\mathcal{N}$ and $\mathcal{P}$ as dynamical variables and using Eq.~\eqref{eq:111}, we obtain the equations of motion for the  population difference and the microscopic polarization.
\begin{equation}\label{GSBE}
\left\{
\begin{array}{l}
\pd{\mathcal{N}(\vp{k},t)}{t}-\frac{e}{\hbar}\vp{E}\cdot\nabla_{\vp{k}}\mathcal{N}(\vp{k},t)=-2\Phi(\vp{k},t)\mathrm{Im}\left\{\mathcal{P}(\vp{k},t)\right\}\\
\\
\pd{\mathcal{P}(\vp{k},t)}{t}-\frac{e}{\hbar}\vp{E}\cdot\nabla_\vp{k}\mathcal{P}(\vp{k},t)=i\varpi_\vp{k}\mathcal{P}(\vp{k},t)
+\frac{i}{2}\Phi(\vp{k},t)\mathcal{N}(\vp{k},t)
\end{array}
\right.
\end{equation}
where $\Phi(\vp{k},t)$ is essentially the matrix element of the external potential for the given Bloch momentum $\vp{k}$ describing the direct optical transition between the upper and lower energy levels. This is due to the fact that the momentum of the light is assumed to be negligibly small.

\begin{equation}
\Phi(\vp{k},t)=\frac{e\vp{E}\cdot\A{\varphi}_{\vp{k}}}{\hbar k}
\end{equation}
The frequency $\hbar\varpi_{\vp{k}}=2\mathcal{E}_{\vp{k}}$ is the energy difference between the upper and lower energy levels. The coupled equations given in Eq.~\eqref{GSBE} are called the semiconductor Bloch equations (SBEs) for graphene. These equations must be solved simultaneously, and subject to the initial condition imposed by the Fermi-Dirac distribution before turning on the field:
\begin{align*}
&\mathcal{N}(\vp{k},-\infty)=\mathcal{N}^{eq}_{\vp{k}}=f(\mathcal{E}(\vp{k}))-f(-\mathcal{E}(\vp{k}))\\
&\mathcal{P}(\vp{k},-\infty)=0
\end{align*}

\section{Perturbative Solution of the Semiconductor Bloch Equations\label{section:5}}
The SBEs describe the quasiclassical transport and interband excitation problems simultaneously. To convert these dynamical equations into a more convenient form,  the transport and interband evolutions are decoupled. The cooperative intraband and interband dynamics can be split by introducing a moving frame in the reciprocal space whose movement is governed by the Boltzmann transport equation \cite{Behrooz}.  The purely classical part of the dynamics drives the distribution function along a trajectory determined by the direction of the electric field.  The time-momentum coordinate in the moving frame is designated by $\{\tau,\vp{k}'\}$. The primed frame is then related to the original frame by $\{\tau,\vp{k}'\}=\{t,\vp{k}-\delta\vp{k}(t)\}$ where $\delta\vp{k}$ obeys 
\begin{equation}\label{eq:Boltzman}
\pd{\delta\vp{k}}{t}+\Gamma\delta\vp{k}=-\frac{e}{\hbar}\vp{E}(t)
\end{equation}
where the extrinsic fitting factor $\Gamma$  accounts for impurity scattering as well as radiation. In the primed frame, the system  evolves  by  pure interband dynamics as described by equations resembling the optical Bloch equations for a generic two-level problem \cite{boyd2003nonlinear}:

\begin{subequations}
\begin{align}
&\pd{\mathcal{N}({\vp{k}'},\tau)}{\tau}=-\gamma^{(1)}_{\vp{k'}}(\mathcal{N}({\vp{k}'},\tau)-\mathcal{N}^{eq}_{\vp{k}'}) 
 -2\Phi(\vp{k}',\tau)\mathrm{Im}\left\{\mathcal{P}(\vp{k}',\tau)\right\}\label{eq:SBE1}\\
&\pd{\mathcal{P}({\vp{k}',\tau)}}{\tau} =-\gamma^{(2)}_{\vp{k}'}\mathcal{P}(\vp{k}',\tau)+
 i\varpi_{\vp{k}'}\mathcal{P}(\vp{k}',\tau)+\frac{i}{2} \Phi(\vp{k}',\tau)\mathcal{N}(\vp{k}',\tau)\label{eq:SBE2}
\end{align}
\end{subequations}
In compliance with the standard notation of Bloch equations, we introduce $w_{\vp{k}}$, $u_{\vp{k}}$, $v_{\vp{k}}$ as 
\begin{subequations}
 \begin{eqnarray}
w_{\vp{k}'}(t) &\triangleq &\mathcal{N}\left(\vp{k}',\tau\right)\label{eq:w}\\
u_{\vp{k}'}(t) &\triangleq & +2\mathrm{Re}\left\{\mathcal{P}\left(\vp{k}',\tau\right)\right\}\label{eq:u}\\
v_{\vp{k}'}(t) & \triangleq &-2\mathrm{Im}\left\{\mathcal{P}\left(\vp{k}',\tau\right)\right\}\label{eq:v}
\end{eqnarray} 
\end{subequations}
where $w_{\vp{k}}$ is the population inversion in the moving frame. The functions $\xi_{\vp{k}}\triangleq {\Phi}(\vp{k}',\tau)$ and $\omega_{\vp{k}} \triangleq \varpi _\vp{k}'$ are also defined for mathematical convenience and are the analytical functions of the exciting field. The function $\xi_{\vp{k}}$ is the equivalent dipole in the moving frame. The optical Bloch equations describe the interband dynamics for the ensemble of the two-level Bloch states. However, due to the motion of the frame, the intraband dynamics are  intertwined with the interband dynamics. The combined nature of the dynamics manifests itself in the time dependence of  index $\vp{k}'$. Since the time scale associated with intraband dynamics is quite different from that of the interband evolution, an adiabatic argument could be applied; however, although the interband transitions  acquire an extra time dependence due to the moving frame, they nevertheless take place independently. The coupled Bloch equations can be converted to the standard optical Bloch equations using the two-level approximation \cite{boyd2003nonlinear}. Henceforth, we drop the prime in the $uvw$-coordinate system:

\begin{subequations}
\begin{align}
&\dot{w}_{\vp{k}}=-\gamma^{(1)}_{\vp{k}}\left(w_{\vp{k}}-w^{eq}_{\vp{k}}\right)+\xi_{\vp{k}} v_{\vp{k}}\label{Wk}\\
&\dot{u}_{\vp{k}}=\omega_{\vp{k}}v_{\vp{k}}-\gamma^{(2)}_{\vp{k}} u_{\vp{k}}\\
&\dot{v}_{\vp{k}}=-\omega_{\vp{k}}u_{\vp{k}}-\gamma^{(2)}_{\vp{k}} v_{\vp{k}}-\xi_{\vp{k}} w_{\vp{k}}
\end{align}   
\end{subequations}
where `dot'  designates the time derivative. The function $w^{eq}_{\vp{k}}$  is the population difference at equilibrium, i.e., $w^{eq}_{\vp{k}}=\mathcal{N}_{\vp{k}}^{eq}$. 
For a weak pump field, the inversion $w_{\vp{k}}$ tends to relax to $w^{eq}_{\vp{k}}$. The coherent terms are the {oscillatory} functions of the field. To proceed further, the current response in the reciprocal space must be identified. According to  Eq.~\eqref{eq:rho100}, together with Eq.~\eqref{eq:J100} the induced current is
\begin{equation}\label{eq:J1000}
\vp{J}=-2ev_F\sum_{\vp{k}}(\A{\vp{k}}\A{\vp{k}}+\A{\varphi}_{\vp{k}}\A{\varphi}_{\vp{k}})\cdot\vec{m}_{\vp{k}}
=ev_F\sum_{\vp{k}}\left[-w_{\vp{k}-\delta\vp{k}}\A{\vp{k}}+v_{\vp{k}-\delta\vp{k}}\A{\varphi}_{\vp{k}}\right]
\end{equation}
Therefore, the equation of motion describing the time evolution of $v_{\vp{k}}$  provides enough information to model the interband response of graphene. Neglecting the time variation of $\omega_{\vp{k}}$ in the adiabatic approximation yields
\begin{equation}
\ddot{v}_{\vp{k}}+2\gamma_2\dot{v}_{\vp{k}}+\left(\omega^2_{\vp{k}}+\gamma_2^2\right)v_{\vp{k}}=-\gamma_2\xi _{\vp{k}}w_{\vp{k}}-\dot{\xi}_{\vp{k}}w_{\vp{k}}-\xi_{\vp{k}}\dot{w}_{\vp{k}} 
\end{equation} 
Since $\omega^2_{\vp{k}}$ is much larger than $\gamma_2^2$ , we can drop $\gamma_2^2 v_{\vp{k}}$ to obtain the result:
\begin{equation}\label{master}
\ddot{v}_{\vp{k}}+2\gamma_2\dot{v}_{\vp{k}}+\omega^2_{\vp{k}}v_{\vp{k}}=-\gamma_2\xi _{\vp{k}}w_{\vp{k}}-\dot{\xi}_{\vp{k}}w_{\vp{k}}-\xi_{\vp{k}}\dot{w}_{\vp{k}} 
\end{equation}
This set of equations can be solved iteratively, leading to a series expansion for $v_{\vp{k}}$ and $w_{\vp{k}}$. The mathematical structure of the coupled damped-driven harmonic oscillator [Eqs.~\eqref{master} and \eqref{Wk}] implies that $v_{\vp{k}}$ contains only odd powers of the field $\xi_{\vp{k}}$ and $w_{\vp{k}}$ contains only even powers of the field:
\begin{subequations}
\begin{align}
&w_{\vp{k}}=w^{eq}_{\vp{k}}+\sum^{\infty}_{n=1}W_{\vp{k}}^{(2n)}\xi_{\vp{k}}^{2n}\label{eq:1000}\\
&v_{\vp{k}}=\sum^{\infty}_{n=1}V_{\vp{k}}^{(2n-1)}\xi_{\vp{k}}^{2n-1}\label{eq:1001}
\end{align}
\end{subequations}
The $n$'th-order expansion terms $w_{\vp{k}}^{(n)}$ and $v_{\vp{k}}^{(n)}$ are defined as
\begin{equation*}
w_{\vp{k}}^{(n)}=W_{\vp{k}}^{(n)}\xi_{\vp{k}}^{n}\quad,\quad v_{\vp{k}}^{(n)}=V_{\vp{k}}^{(n)}\xi_{\vp{k}}^{n} 
\end{equation*}
The iterative procedure can be conveniently carried out by defining  the operators  $\bm{\mathcal{V}}_{\vp{k}}$ and $\bm{\mathcal{W}}_{\vp{k}}$. 
\begin{subequations}
\begin{align}
\bm{\mathcal{V}}_{\vp{k}}(\omega) \triangleq & \frac{\gamma^{(2)}_\vp{k}+i\omega}{\omega^2-2i\gamma^{(2)}_\vp{k}\omega-\varpi^2_\vp{k}}\frac{e}{\hbar k}\A{\varphi}_{\vp{k}}\\
\bm{\mathcal{W}}_{\vp{k}}(\omega) \triangleq &\frac{1}{i\omega+\gamma^{(1)}_\vp{k}}\frac{e}{\hbar k}\A{\varphi}_{\vp{k}}
\end{align}
\end{subequations}
 
In addition to the purely interband multi-photon  process described above, a part of the nonlinearity originates from the quasiclassical transport or intraband transitions. 
As will be clarified further in subsequent sections, the frequency mixing effects in graphene arise from the pure intraband, pure interband and interband-intraband transitions. The pure intraband response occurs only due to the change in  the population difference. Using Eq.~\eqref{eq:J1000}, the intraband contribution to the current is
\begin{equation}\label{eq:10300}
\vp{J}_{intra}
=-ev_F\sum_{\vp{k}}w^{eq}_{\vp{k}-\delta\vp{k}}\A{\vp{k}}
\end{equation}
then, the $n$'th-order nonlinearity due to the \textit{pure intraband process} can be obtained using the following Taylor expansion:
\begin{equation}\label{Master-Intraband1}
\vp{J}^{(n)}_{intra}=(-1)^{n+1}\frac{ev_F}{n!}\sum_{\vp{k}}\A{\vp{k}}\left[\delta\vp{k}\cdot\nabla_{\vp{k}}\right]^n \mathcal{N}^{eq}_{\vp{k}}
\end{equation}
For the sake of mathematical convenience, the derivative operator is represented as 
\begin{equation}\label{eq:D}
\A{\bm{\mathcal{D}}}_{\vp{k}}(\omega) \triangleq \frac{1}{i\omega+\Gamma}\frac{e}{\hbar}\nabla_{\vp{k}}
\end{equation}
where the Drude-like coefficient $1/(i\omega+\Gamma)$ is obtained from Eq.~\eqref{eq:Boltzman}. An intuitive symmetry argument shows that in graphene---a centrosymmetric crystal---even orders of nonlinearity do not exist. The derivations of the linear and third-order conductivity tensors of graphene are discussed below.

\subsection{Linear Optical Response of Graphene}

Equation~\eqref{eq:10300} for $n=1$ gives the intraband conductivity tensor in the $k$-space
\begin{equation}
\OO{\sigma}^{(1)}_{intra}(\vp{k},\omega_p)=-ev_F\A{\vp{k}}\A{\bm{\mathcal{D}}}_{\vp{k}}(\omega_p)\mathcal{N}^{eq}_{\vp{k}}
\end{equation}
where we have assumed that the electric field is $\vp{E}(t)=\T{\vp{E}}_p\exp(i\omega_p t)+c.c $ and $ \OO{\sigma}^{(1)}_{intra}({\vp{k}},\omega_p)$ is defined via 
\begin{equation*}
\vp{J}_{\vp{k}}=\OO{\sigma}^{(1)}_{intra}({\vp{k}},\omega_p)\cdot\vp{E}
\end{equation*}
Performing the integration in the reciprocal space, the off-diagonal terms vanish and we arrive at
\begin{equation}
\sigma^{(1)}_{intra}(\omega_p)=\frac{g_sg_v}{4\pi}\frac{e^2}{\hbar^2}\frac{1}{\left(i\omega_p+\Gamma\right)}
\int_0^{+\infty}\mathrm{d}\mathcal{E}\mathcal{E}\left[\pd{f(\mathcal{E})}{\mathcal{E}}-\pd{f(-\mathcal{E})}{\mathcal{E}}\right]
\end{equation}
where $g_s$ and $g_v$ are the spin and valley degeneracy factors, respectively.

The interband linear optical response of graphene can be obtained using the master equation \eqref{master}. The linear optical response is a single-photon process and unlike higher-order terms it can be  obtained {independently}  for inter- and intraband contributions. The first-order solution of  Eq.~\eqref{master} can be derived by replacing $w_{\vp{k}}$ and $\dot{w}_{\vp{k}}$  with $w^{eq}_{\vp{k}}$ and $0$, respectively,
\begin{equation}
\OO{\sigma}_{inter}(\vp{k},\omega_p)=ev_F\A{\varphi}_{\vp{k}}\bm{\mathcal{V}}_{\vp{k}}(\omega_p)\mathcal{N}^{eq}_{\vp{k}}
\end{equation} 
Integration over the reciprocal space and including the density of states gives
\begin{equation}\label{eq:interband-linear}
\sigma^{(1)}_{inter}(\omega_p)=\frac{e^2}{\hbar}\frac{g_sg_v}{4\pi}
\int_0^{+\infty}\mathrm{d}\mathcal{E}
\frac{\left(\gamma_2+i\omega_p\right)}{\omega_p^2-2i \gamma_2\omega_p-\varpi_{\vp{k}}^2}\left[f(\mathcal{E})-f(-\mathcal{E})\right]
\end{equation}

\subsection{Third Order Frequency Mixing in Graphene}
\begin{figure}[t]
\centering
\includegraphics[scale=0.04]{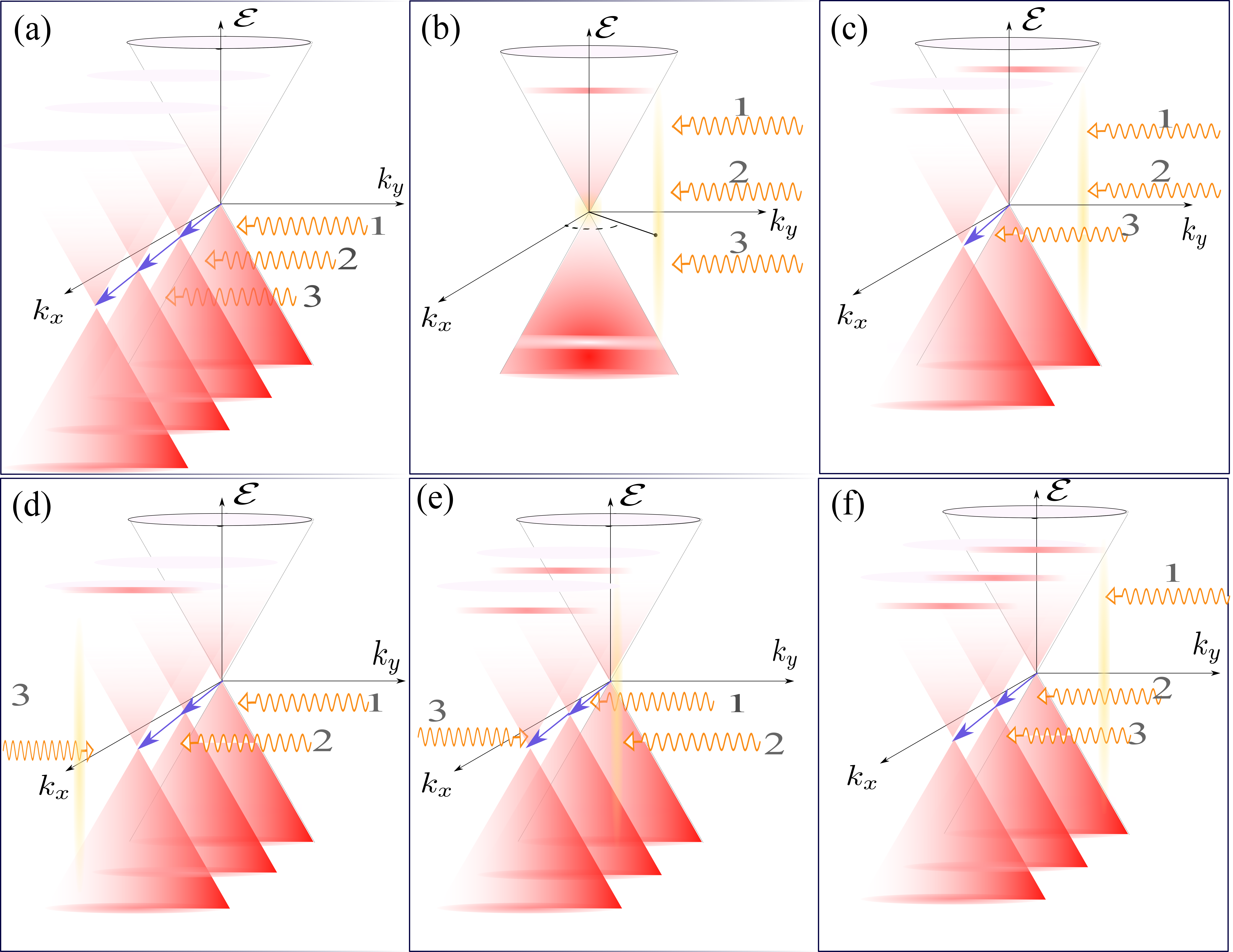}
\caption{\small Schematic representation of different three-photon processes contributing to the third-order nonlinear optics of graphene. The intraband type of dynamics is depicted by displacement of the distribution and the interband dynamics are shown by two-level transitions.}
\label{fig:trasnistions}
\end{figure}

Throughout this section we assume that three complex fields with a time dependence of $e^{i\omega_pt}$, $e^{i\omega_qt}$ and $e^{i\omega_rt}$ are mixing through the third-order conductivity of graphene. As mentioned earlier, the third-order optical response can be interpreted as a three-photon process and  different terms contribute to the third-order conductivity tensor, namely a  pure intraband term, a pure interband term, and a combination of both.
The distinct photon processes contributing to the third-order optics of graphene are schematically shown in Fig.~\ref{fig:trasnistions}. The intraband dynamics causes the quasiparticles to travel along the trajectory determined by the direction of the electric field at the graphene plane. The quasiclassical Boltzman-like dynamics are pictured schematically by displacement of the entire distribution of the quasiparticles over the reciprocal space. The interband contributions are  shown by two-level transitions of the quasiparticles predominantly around the zero detuning region.  The adopted mathematical structure allows us to easily find the conductivity tensors associated with the six processes shown in Fig.~\ref{fig:trasnistions}.
 
\subsubsection{Pure intraband third-order nonlinearity}
The pure intraband contribution is fundamentally part of the third-order nonlinearity which  occurs exclusively due to the Boltzman-like transport. The third-order intraband process is schematically displayed in Fig.~\ref{fig:trasnistions}{a}. 
Equation \eqref{Master-Intraband1} determines the nonlinear contribution of the third-order intraband evolution as 

\begin{equation}
\overline{\overline{\mathcal{I}}}^{(3)}_{\vp{k},1}(\omega_p,\omega_q,\omega_r)=-ev_F\mathcal{P}_I\left\{ \A{\vp{k}}\A{\bm{\mathcal{D}}}_{\vp{k}}(\omega_r)\A{\bm{\mathcal{D}}}_{\vp{k}}(\omega_q)\A{\bm{\mathcal{D}}}_{\vp{k}}(\omega_p)\mathcal{N}^{eq}_{\vp{k}} \right\}
\end{equation} 
where the normalized gradient operator $\A{\bm{\mathcal{D}}}_{\vp{k}}$ was introduced in Eq.~\eqref{eq:D}. The subscript `$1$' is used to identify the intraband contribution and the superscript  `$(3)$' refers to the third-order effect. Here we have made use of the intrinsic permutation operator $\mathcal{P}_I$;  all possible  
permutations of the input frequencies $\omega_p$, $\omega_q$, and $\omega_r$ must be summed over. The overall intraband conductivity tensor is then obtained as 
\begin{equation}
\overline{\overline{\sigma}}^{(3)}_{intra}(\omega_p,\omega_q,\omega_r)=\sum_{\vp{k}}\overline{\overline{\mathcal{I}}}^{(3)}_{\vp{k},1}
\end{equation}

\subsubsection{Pure interband third-order nonlinearity}
Pure interband third-order nonlinearity can be obtained using the master equations \eqref{master} and \eqref{Wk} in the moving frame. Power expansion of the inversion and the polarization  in terms of the exciting field as in Eqs.~\eqref{eq:1000} and \eqref{eq:1001} gives $v^{(3)}_{\vp{k}}$. Assuming that the first photon $\omega_p$ provides time variation for $v^{(1)}_{\vp{k}}(\omega_p)$, the first nonzero oscillatory component of  $w_{\vp{k}}$ as well as the third harmonic of $v^{(3)}_{\vp{k}}$ can be found from the following equations 
\begin{eqnarray}
\dot{w}^{(2)}_{\vp{k}}+\gamma_1 w_{\vp{k}}^{(2)} &= &\xi_{\vp{k}}(\omega_q)v^{(1)}_{\vp{k}}(\omega_p)\\
\ddot{v}^{(3)}_{\vp{k}}+2\gamma_2\dot{v}^{(3)}_{\vp{k}}+\omega^2_{\vp{k}}v^{(3)}_{\vp{k}} &=&
 -\gamma_2\xi _{\vp{k}}(\omega_r)w^{(2)}_{\vp{k}}-\dot{\xi}_{\vp{k}}(\omega_r)w^{(2)}_{\vp{k}}-\xi_{\vp{k}}(\omega_r)\dot{w}^{(2)}_{\vp{k}} 
\end{eqnarray}  
These equations can be solved simultaneously to find the contribution of the interband dynamics associated with Bloch index $\vp{k}$ to the third-order optical nonlinearity. Making use of the operators $\bm{\mathcal{V}}_{\vp{k}}$ and $\bm{\mathcal{W}}_{\vp{k}}$  allows a compact solution as
\begin{equation}
\overline{\overline{\mathcal{I}}}^{(3)}_{\vp{k},2}(\omega_p,\omega_q,\omega_r)=-ev_F\mathcal{P}_I\left\{  
\A{\varphi}_{\vp{k}}\bm{\mathcal{V}}_{\vp{k}}(\omega_p+\omega_q+\omega_r)\bm{\mathcal{W}}_{\vp{k}}(\omega_p+\omega_q) \bm{\mathcal{V}}_{\vp{k}}(\omega_p)\mathcal{N}^{eq}_{\vp{k}} \right\}
\end{equation}
Performing the integration in the reciprocal space and applying the permutation operator yields the final expression for the interband conductivity tensor. The interband conductivity tensor is then obtained as
\begin{equation}
\overline{\overline{\sigma}}^{(3)}_{inter}(\omega_p,\omega_q,\omega_r)=\sum_{\vp{k}}\overline{\overline{\mathcal{I}}}^{(3)}_{\vp{k},2}
\end{equation}
where the summation goes over all quantum states. This term is obviously singular at the origin.

\subsubsection{Interband-intraband third-order nonlinearity}
The master equations together with the notion of a moving frame lead to four possible combinations of the inter- and intraband dynamics shown in Fig.~\ref{fig:trasnistions}{c-f}. 
\begin{itemize}
\item 
\textbf{Interband-Interband-Intraband:} According to Fig.~\ref{fig:trasnistions}{c}, the interband dynamics modify the population difference through a non-parametric transition. The third photon (which is actually the probing photon!)\ causes an intraband transition. The master equations \eqref{master} and \eqref{Wk} yield the modified population accomplished via two subsequent photon processes.  
\begin{equation*}
\dot{w}^{(2)}_{\vp{k}}+\gamma_1 w_{\vp{k}}^{(2)} = \xi_{\vp{k}}(\omega_q)v^{(1)}_{\vp{k}}(\omega_p)
\end{equation*}
The third process causes the modified population to move over the reciprocal space and creates an intraband current described by equation \eqref{eq:10300}. The tensor associated with this process reads

\begin{equation}
\overline{\overline{\mathcal{I}}}^{(3)}_{\vp{k},3}(\omega_p,\omega_q,\omega_r)=-ev_F\mathcal{P}_I\left\{ \A{\vp{k}}\A{\bm{\mathcal{D}}}_{\vp{k}}(\omega_r) \bm{\mathcal{W}}_{\vp{k}}(\omega_p+\omega_q) \bm{\mathcal{V}}_{\vp{k}}(\omega_p)\mathcal{N}^{eq}_{\vp{k}} \right\}
\end{equation}

\item
\textbf{Intraband-Intraband-Interband:} Following two subsequent intraband transitions at the frequencies $\omega_p$ and $\omega_q$, the population difference oscillates at the sum frequency $\omega_p+\omega_q$. The third photon probes the graphene whose quasiparticles are oscillating over the reciprocal space and the current is induced due to interband transition. This process is sketched in Fig.~\ref{fig:trasnistions}{d} and is mathematically encoded by $\overline{\overline{\mathcal{I}}}^{(3)}_{\vp{k},4}$. 
\begin{equation}
\overline{\overline{\mathcal{I}}}^{(3)}_{\vp{k},4}(\omega_p,\omega_q,\omega_r)=-ev_F\mathcal{P}_I\left\{  
\A{\varphi}_{\vp{k}}\bm{\mathcal{V}}_{\vp{k}}(\omega_p+\omega_q+\omega_r)\A{\bm{\mathcal{D}}}_{\vp{k}}(\omega_q)\A{\bm{\mathcal{D}}}_{\vp{k}}(\omega_p)\mathcal{N}^{eq}_{\vp{k}} \right\}
\end{equation}

\item
\textbf{Intraband-Interband-Intraband:} Fig.~\ref{fig:trasnistions}{e} displays the ordering of the processes. Following an intraband transition, the population difference oscillates at frequency~$\omega_p$. The second process is interband  which creates a coherence (polarization) at the frequency of $\omega_p+\omega_q$. The induced polarization is driven by  an additional intraband transition to create a current at the frequency of $\omega_p+\omega_q+\omega_r$. Although the last transition is intraband, the induced current is of interband nature and is modulated by the moving frame.

\begin{equation}
\overline{\overline{\mathcal{I}}}^{(3)}_{\vp{k},5}(\omega_p,\omega_q,\omega_r)=-ev_F\mathcal{P}_I\left\{ 
\A{\varphi}_{\vp{k}}\A{\bm{\mathcal{D}}}_{\vp{k}}(\omega_r)\bm{\mathcal{V}}_{\vp{k}}(\omega_p+\omega_q)\A{\bm{\mathcal{D}}}_{\vp{k}}(\omega_p)\mathcal{N}^{eq}_{\vp{k}} \right\}
\end{equation}

\item
\textbf{Interband-Intraband-Intraband:} An interband transition caused by the first photon creates the polarization oscillating at frequency $\omega_p$. Due to the Boltzman-like transport, the induced polarization is modulated by two consecutive harmonics $\omega_q$ and $\omega_r$. The induced current has an interband nature and is shown in Fig.~\ref{fig:trasnistions}{f}. The overall dynamics are encoded by $\overline{\overline{\mathcal{I}}}^{(3)}_{\vp{k},6}$:
\begin{equation}
\overline{\overline{\mathcal{I}}}^{(3)}_{\vp{k},6}(\omega_p,\omega_q,\omega_r)=-ev_F\mathcal{P}_I\left\{
\A{\varphi}_{\vp{k}}\A{\bm{\mathcal{D}}}_{\vp{k}}(\omega_r)\A{\bm{\mathcal{D}}}_{\vp{k}}(\omega_q)\bm{\mathcal{V}}_{\vp{k}}(\omega_p)\mathcal{N}^{eq}_{\vp{k}} \right\}
\end{equation}

\end{itemize}

The overall intraband-interband conductivity tensor is then obtained as 
\begin{equation}
\overline{\overline{\sigma}}^{(3)}_{intra-inter}(\omega_p,\omega_q,\omega_r)=\sum_{l=3}^{6}\sum_{\vp{k}}\overline{\overline{\mathcal{I}}}^{(3)}_{\vp{k},l}
\end{equation}

\section{Length Gauge versus Velocity Gauge in Graphene\label{app:gauge}}

In the long-wavelength regime where the spatial dependence of the electromagnetic field can be neglected, there are two theoretical approaches to couple electromagnetic radiation into the Hamiltonian. Within the so-called \textit{velocity gauge}, light and matter are coupled via the minimal substitution of the vector potential $\vp{A}(t)$, with the electric field given by $\vp{E}(t)=-\partial\vp{A}/\partial{t}$. The alternative approach is referred to as the \textit{length gauge}~\cite{aversa1995nonlinear} where the field is directly coupled by means of the additive scalar potential $V(\vp{r})=e\vp{E}\cdot\vp{r}$ with the elementary charge $e>0$. Although the velocity gauge preserves translation symmetry and different Bloch states remain uncoupled, there are several undesirable features associated with the velocity gauge that plague the calculations~\cite{aversa1995nonlinear,Virk2007}. 

In particular, the  treatment of the nonlinear optical response within the velocity gauge is  susceptible to numerical errors caused by truncation of the band space \cite{aversa1995nonlinear,Ventura2017,PhysRevB.96.195413}. Since the electric field is proportional to the time derivative of the vector potential, poles at $\omega=0$ inevitably appear.  Diverging terms are not obviously real in semiconductors. It has been rigorously proven that they essentially vanish due to time reversal symmetry and an effective mass sum rule \cite{aversa1995nonlinear}. However, a full calculation of the optical transitions over the entire band is required to eliminate the diverging terms \cite{Virk2007} and numerical errors due to truncation amplify. Moreover, in two-level systems where an effective Hamiltonian is intended to describe the electrons' dynamics (as in the case of graphene), the application of  the velocity gauge is quite vulnerable \cite{Ventura2017}. References \cite{Ventura2017} and \cite{aversa1995nonlinear} show that in the perturbative treatment of the nonlinear optical response of such systems, the contribution made by the remaining parts of the band cannot be ignored in the calculations. This observation  implies that a \textit{local interpretation} of the optical transitions within the velocity gauge is not reliable and that all  Bloch states collectively impact the optical response \cite{Ventura2017}.  

To cope with this situation, the length gauge can be employed. The different contributions of the position operators have been extensively discussed in the literature \cite{aversa1995nonlinear,Virk2007} and it has been shown that the perturbative treatment of the nonlinear optical response of semiconductors can be reliably performed in a two-level model by making use of the length gauge \cite{AlNaib2014}. This gauge eliminates the nonphysical diverging terms and renders it possible to  interpret the optical transition locally over the band space. The matrix element of the position operator between the Bloch states indexed by $(\vp{k},s)$ and $(\vp{k}',s')$ is \cite{Virk2007}
\begin{equation}\label{eq:r-app}
\vp{r}_{\vp{k}s,\vp{k}'s'}=\delta(\vp{k}-\vp{k}')[i\delta_{ss'}\pd{}{\vp{k}}+i\boldsymbol{\zeta}_{ss'}(\vp{k})]
\end{equation}
The position operator can be interpreted as the generator of  translation in the space of the Bloch functions \cite{Virk2007}. The Berry potential $\boldsymbol{\zeta}_{ss'}(\vp{k})$ is required as the geometric phase correction and, therefore, its action in nontrivial topologies is crucial. 

To shed light on the applicability of the velocity gauge to perturbation theory, as a benchmark, we proceed with finding the linear optical response of graphene within the velocity gauge.  Assume that a graphene layer lying in the $xy$ plane is illuminated by an obliquely incident electromagnetic field $\vp{E}(\vp{r})=\vp{E}_0\exp\left(i\omega t-i\vp{k}_0.\vp{r}\right)$, where $\vp{k}_0$ is the wavenumber of the incident field. For mathematical convenience, the tangential component of $\vp{k}_0$ is denoted by $\vp{q}\triangleq\vp{k}_0\cdot(\A{x}\A{x}+\A{y}\A{y})$. Within the velocity gauge, the time derivative of the divergence-less vector potential $\vp{A}$ is linearly related to the electric field. The  vector potential is
\begin{equation}\label{eq:A-E}
-i\omega\vp{A}=\vp{E}(\vp{r})
\end{equation}
The interaction Hamiltonian is then $\A{H}_I=\vp{A}\cdot\A{\vec{\mathcal{J}}}$ where $\A{\vec{\mathcal{J}}}$ is the current operator derived in Eq.~\eqref{eq:J100}. Using the customary minimal substitution prescription, the overal Hamiltonian reads 
\begin{equation}
\A{H}=\A{H}_0+\vp{A}\cdot\A{\vec{\mathcal{J}}}=\A{H}_0+\vp{A}_0\cdot\A{\vec{\mathcal{R}}}(-\vp{q})
\end{equation}
For compactness, we have defined the operator $\A{\vec{\mathcal{R}}}(\vp{q})$ as 
\begin{equation}
\A{\vec{\mathcal{R}}}(\vp{q})=\A{\vec{\mathcal{J}}}\exp(+i\vp{q}\cdot\bm{\rho})
\end{equation}
The linear variation of the density matrix due to the presence of the external potential is calculated using the perturbation expansion
\begin{equation}
\A{\rho}=\A{\rho}_0+\delta\A{\rho}^{(1)}+\cdots
\end{equation}
The first-order perturbation theory gives $\delta\A{\rho}^{(1)}$, which linearly depends on the electric field:
\begin{equation}
\delta\A{\rho}^{(1)}_{\vp{k}s,\vp{k}'s'}=\frac{f({E_{\vp{k}',s'}})-f({E_{\vp{k},s}})}{E_{\vp{k}',s'}-E_{\vp{k},s}+\hbar\omega+i\gamma_{ss'}}
\bra{\vp{k},s}\A{H}_{I}\ket{\vp{k}',s'}
\end{equation}
where $f({E})$ is the Fermi-Dirac distribution.The index $s$ refers to the upper and lower energy states. The parameter $\gamma$ appearing in the denominator is the phenomenological relaxation coefficient. The first-order induced current is then  obtained as 
\begin{equation}
\vp{J}_\vp{q}=\mathrm{Tr}\left\{\delta\A{\rho}^{(1)}\A{\vec{\mathcal{R}}}(\vp{q})\right\}
\end{equation}
\begin{equation}\label{eq:J_velocity}
\vp{J}_\vp{q}=\vp{A}_0\cdot\sum_{ss'}\frac{f(E_{\vp{k}',s'})-f(E_{\vp{k},s})}{E_{\vp{k}',s'}-E_{\vp{k},s}+\hbar\omega+i\gamma_{ss'}}\bra{s}\A{\vec{\mathcal{R}}}_{\vp{k}s,\vp{k}'s'}(-\vp{q})\ket{s'}\bra{s'}\A{\vec{\mathcal{R}}}_{\vp{k}'s',\vp{k}s}(\vp{q})\ket{s}
\end{equation}
where 
\begin{equation}
\A{\vec{\mathcal{R}}}_{\vp{k}s,\vp{k}'s'}(\vp{q})\approx\left(\A{\vp{u}}\sigma_z+\A{\bm{\varphi}}_u\sigma_y\right)(-ev_F)
\end{equation}
From now on, we define $\Pi_{\vp{k}\vp{k}'}^{ss'}$ as
\begin{equation}
\Pi_{\vp{k}\vp{k}'}^{ss'}\triangleq\frac{f({E_{\vp{k}',s'}})-f({E_{\vp{k},s}})}{E_{\vp{k}',s'}-E_{\vp{k},s}+\hbar\omega+i\gamma_{ss'}}
\end{equation}
The unit vectors $\A{\vp{u}}$ and $\A{\bm{\varphi}}_u$ are defined as
\begin{equation}
\A{\vp{u}}(\vp{k},\vp{q})=\frac{\A{\vp{k}}'(\vp{k},\vp{q})+\A{\vp{k}}}{\abs{\A{\vp{k}}'(\vp{k},\vp{q})+\A{\vp{k}}}}\quad ,\quad \A{\bm{\varphi}}_u(\vp{k},\vp{q})=\A{z}\times\A{\vp{u}}(\vp{k},\vp{q})
\end{equation}
where $\A{\vp{k}}'(\vp{k},\vp{q})=(\vp{k}+\vp{q})/\abs{\vp{k}+\vp{q}}$. The integrand in Eq.~\eqref{eq:J_velocity} is explicitly written as $\OO{\xi}_{\vp{k},\vp{q}}$ defined as
\begin{equation}
\OO{\xi}_{\vp{k},\vp{q}}(\omega)
\triangleq
\left(\Pi_{\vp{k},\vp{k}+\vp{q}}^{11}+\Pi_{\vp{k},\vp{k}+\vp{q}}^{22}\right)\A{\vp{u}}(\vp{k},-\vp{q})\A{\vp{u}}(\vp{k},\vp{q})+\left(\Pi_{\vp{k},\vp{k}+\vp{q}}^{12}+\Pi_{\vp{k},\vp{k}+\vp{q}}^{21}\right)\A{\bm{\varphi}}(\vp{k},-\vp{q})\A{\bm{\varphi}}(\vp{k},\vp{q})
\end{equation}

Using Eq.~\eqref{eq:A-E}, the conductivity tensor associated with the transitions between the quasiparticles with Bloch indices $\vp{k}$ and $\vp{k}+\vp{q}$ is then obtained as $\OO{\sigma}^{(1)}_{\vp{k},\vp{q}}=i\OO{\xi}_{\vp{k},\vp{q}}(\omega)/\omega$. However, as we expected,  this term is diverging at lower frequencies due to the pole at $\omega=0$. It is at this point that the undesirable features of the velocity gauge manifest themselves. The diverging term appears in the optical response and, therefore, special care must be taken. As discussed, this singularity would be resolved in a full-band calculation where all contributions are included. In order to eliminate this singularity, one can manually add a zero to the optical response to cancel out the pole at $\omega=0$ and  yield a physically correct result:
\begin{equation}
\tilde{\OO{\sigma}}^{(1)}_{\vp{k},\vp{q}}=\frac{i}{\omega}\left[\OO{\xi}_{\vp{k},\vp{q}}(\omega)-\OO{\xi}_{\vp{k},\vp{q}}(0)\right]
\end{equation} 
Accordingly, we define $\tilde{\Pi}_{\vp{k}\vp{k}'}^{ss'}$ as 
\begin{equation}
\tilde{\Pi}_{\vp{k}\vp{k}'}^{ss'}\approx-\frac{1}{E_{\vp{k}',s'}-E_{\vp{k},s}}\frac{f({E_{\vp{k}',s'}})-f({E_{\vp{k},s}})}{E_{\vp{k}',s'}-E_{\vp{k},s}+\hbar\omega+i\gamma_{ss'}}
\end{equation}
which yields
\begin{equation}\label{eq:sigmak_velcoity}
\tilde{\OO{\sigma}}_{\vp{k},\vp{q}}(\omega)
\triangleq
\left(\tilde{\Pi}_{\vp{k},\vp{k}+\vp{q}}^{11}+\tilde{\Pi}_{\vp{k},\vp{k}+\vp{q}}^{22}\right)\A{\vp{u}}(\vp{k},-\vp{q})\A{\vp{u}}(\vp{k},\vp{q})+\left(\tilde{\Pi}_{\vp{k},\vp{k}+\vp{q}}^{12}+\tilde{\Pi}_{\vp{k},\vp{k}+\vp{q}}^{21}\right)\A{\bm{\varphi}}(\vp{k},-\vp{q})\A{\bm{\varphi}}(\vp{k},\vp{q})
\end{equation}
The conductivity tensor now reads
\begin{equation}
\OO{\sigma}^{(1)}(\omega,\vp{q})=i\frac{e^2}{\hbar}g_sg_v\frac{1}{4\pi^2}\iint\mathrm{d}\mathcal{E}_x\mathrm{d}\mathcal{E}_y\tilde{\OO{\sigma}}_{\vp{k},\vp{q}}(\omega)
\end{equation}
where $\mathcal{E}_i=\hbar v_F k_i$ is defined to make the integral dimensionless. In the long-wavelength limit, the optical conductivity obtained 
above and the results of the calculations within the length gauge offer identical expressions for the linear conductivity tensor. It should be noted that 
\begin{subequations}
\begin{align}
&\lim_{q\to 0} \tilde{\Pi}_{\vp{k},\vp{k}+\vp{q}}^{ss}=\pd{f(\mathcal{E}_{s,k})}{\mathcal{E}} \frac{1}{\omega+i\Gamma}\\
& \lim_{q\to 0} \A{\vp{u}}(\vp{k},-\vp{q})= \lim_{q\to 0} \A{\vp{u}}(\vp{k},\vp{q})=\A{\vp{k}}
\end{align}
\end{subequations}
where $\gamma_{ss}=\Gamma$ is, by definition, the intraband relaxation coefficient.  This part of the conductivity is  obviously responsible for the intraband dynamics  manifested in the $\partial/\partial\vp{k}$ terms appearing in Eq.~\eqref{eq:r-app}. Likewise, for the interband contribution where $s\neq s'$
\begin{subequations}
\begin{align}
&\lim_{q\to 0} \tilde{\Pi}_{\vp{k},\vp{k}+\vp{q}}^{ss'}=\pm\frac{1}{2\mathcal{E}_{s,k}}\frac{f({\mathcal{E}_{k,s'}})-f({\mathcal{E}_{k,s}})}{\mathcal{E}_{k,s'}-\mathcal{E}_{k,s}+\hbar\omega+i\gamma_{\vp{k}}}\\
& \lim_{q\to 0}\A{\bm{\varphi}}(\vp{k},-\vp{q})=\lim_{q\to 0} \A{\bm{\varphi}}(\vp{k},\vp{q}) = \A{\bm{\varphi}}_{\vp{k}}
\end{align}
\end{subequations}
where $\mathcal{E}_{k,s}=-\mathcal{E}_{k,s'}$. The prefactor $1/{2\mathcal{E}_{s,k}}$  together with the multiplicative vector $\A{\bm{\varphi}}_{\vp{k}}$ corresponds to $\boldsymbol{\zeta}_{ss'}(\vp{k})$ appearing in  Eq.~\eqref{eq:r-app}. 

In conclusion, by removing the artificial diverging pole in the velocity gauge, both approaches yield identical results. The aforementioned pole arises due to the band space truncation and it can be removed by developing the sum rule reported in Ref.~\onlinecite{aversa1995nonlinear}. In the effective two-band model, the velocity gauge should be repaired to account for the nonphysical terms and as a result, the perturbative treatment of the optical response of graphene within the velocity gauge, in its original form, is not perfectly reliable. The length gauge can be consistently employed to develop a full theoretical model for the nonlinear optics of graphene.  

\newpage

\section{Relaxation Dynamics and Phenomenological Coefficients\label{app:Relaxation}}

The adopted theoretical models for graphene in this paper are based on a single-particle picture that treats many-body effects at the phenomenological level. In some  circumstances, a more accurate inclusion of the relaxation coefficients would provide a substantial improvement in understanding  the underlying physics. Striving to use a single-particle description of the electron dynamics in graphene, the many-body effects are analyzed within a microscopic theory (detailed below) and the microscopic relaxation coefficients (k-dependent coefficients) are obtained numerically by means of a long pulse excitation. Inspired by the practical methods of measurement in the laboratory, we excite graphene within the many-body model by a long and sharp edged pulse to resolve the  relaxation time  for the microscopic polarization $\mathcal{P}({\vp{k}},t)$ and the population $\mathcal{N}(\vp{k},t)$.

\subsection{Model Description}
We have utilized the microscopic  model developed  in Ref.~\cite{Malic2011}, which takes into account the  Coulomb and phonon-induced relaxation channels. The adopted model employs a many-body Hamiltonian that consists of a free-carriers (electron and phonon) part $H_0$, the carrier field $H_{c-f}$, and the carrier-carrier and carrier-phonon Hamiltonians which are represented by $H_{c-c}$ and $H_{c-p}$ respectively \cite{Malic2011}:
\begin{equation}
H=H_0+H_{c-f}+H_{c-c}+H_{c-p}
\end{equation}
The approximations in this model are concisely outlined as follows:
\begin{enumerate}[(I)]
\item
The free-carriers (band induced) Hamiltonian $H_0$ consists of free-electron and free-phonon  dynamics. The free-electron Hamiltonian is constructed based on a full-band model beyond the Dirac cone approximation for weakly bounded $2p_z$ orbitals. The full-band model is capable of resolving exciting effects in the absorption spectrum of graphene. More importantly, phonon-induced intervalley scattering is efficiently modeled in the full-band Hamiltonian.   The Hamiltonian $H_0$ also contains the dispersion of the free phonons including both the optical and acoustic ones. The optical phonons possess two sharp kinks in their dispersion around the high-symmetry points \cite{Piscanec2004}. The corresponding phonon modes causes strong electron-phonon coupling, which in turn yields nonzero population relaxation around the Dirac points \cite{Malic2011}.
\item
 The carrier-field coupling ($H_{c-f}$) is obtained using minimal coupling into the conical momentum. Since we are using a full-band model, the velocity gauge offers reliable results. 
 \item
The electron-electron interaction has been modeled within the first-order mean-field Hartree-Fock approximation. The electrostatic screening effect is calculated using an effective single-particle picture that leads to the Lindhard approximation of the dielectric function. It is worth pointing out  that the calculations are carried out in the static limit. The low-energy excitations (particularly the ones in the vicinity of the Dirac points) are minimally affected by this approximation.  Many-body dephasing due to radiation is almost nonexistent in the low-frequency limit due to the reduced phase space accessible to the low-energy photons.  
\item
 The electron-phonon coupling ($H_{c-p}$) is also obtained accurately using phonon-induced deformation potentials. The appropriate potentials for the optical and acoustic phonons are calculated in Ref.~\onlinecite{Piscanec2004}.
\end{enumerate}

\subsection{Effective Relaxation Coefficients}
The numerical estimation of the microscopic relaxation dynamics is carried out by \textit{projecting} the many-body dynamics into the SBEs obtained from the reduced Liouville's equation. In compliance with the mathematical structure of the SBEs, we excite the graphene Bloch equations (full microscopic dynamics) with a relatively long and sharp-edged pulse and capture the relaxation time scale. The long duration of the optical excitation  allows us to capture the steady-state dynamics and the sharp edged nature of the excitation facilitates time-domain characterization of the microscopic relaxation coefficients. We then  curve-fit the envelope of the microscopic polarization and population (after the pulse is tuned off) to an exponential decay. The numerical steps are outlined below.

\begin{figure}[t]
\centering
\includegraphics[width=0.85\linewidth]{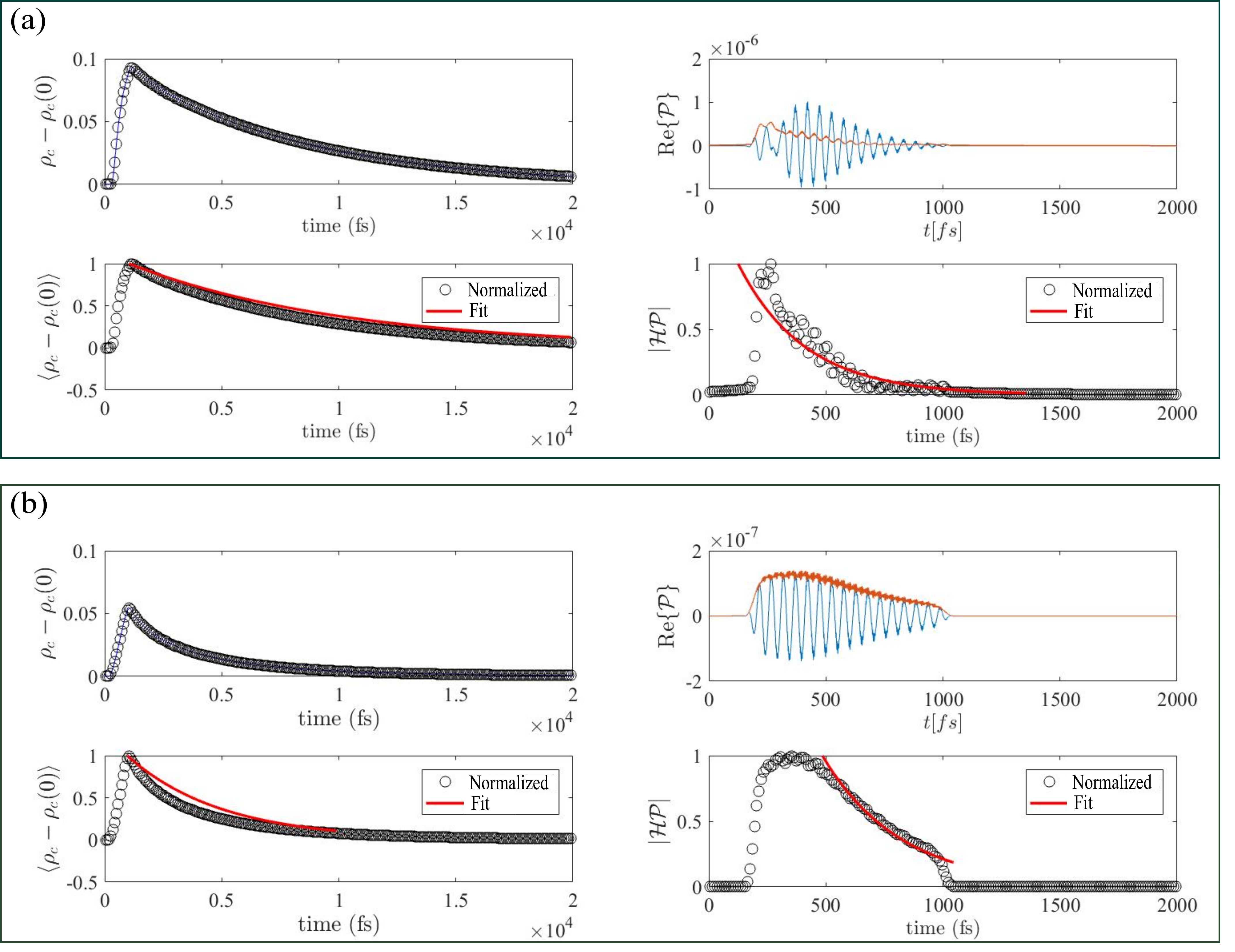}
\caption{\small Extraction of the microscopic relaxation coefficients for the optical excitation of $\hbar\omega=80\mathrm{meV}$ at (a) $\hbar v_F k=10\mathrm{meV}$(low energy) and (b) for $\hbar v_F k=100\mathrm{meV}$ (high energy) points.}
\label{fig:extraction_gamma-80meV}
\end{figure}

\begin{figure}[b]
\centering
\includegraphics[width=0.85\linewidth]{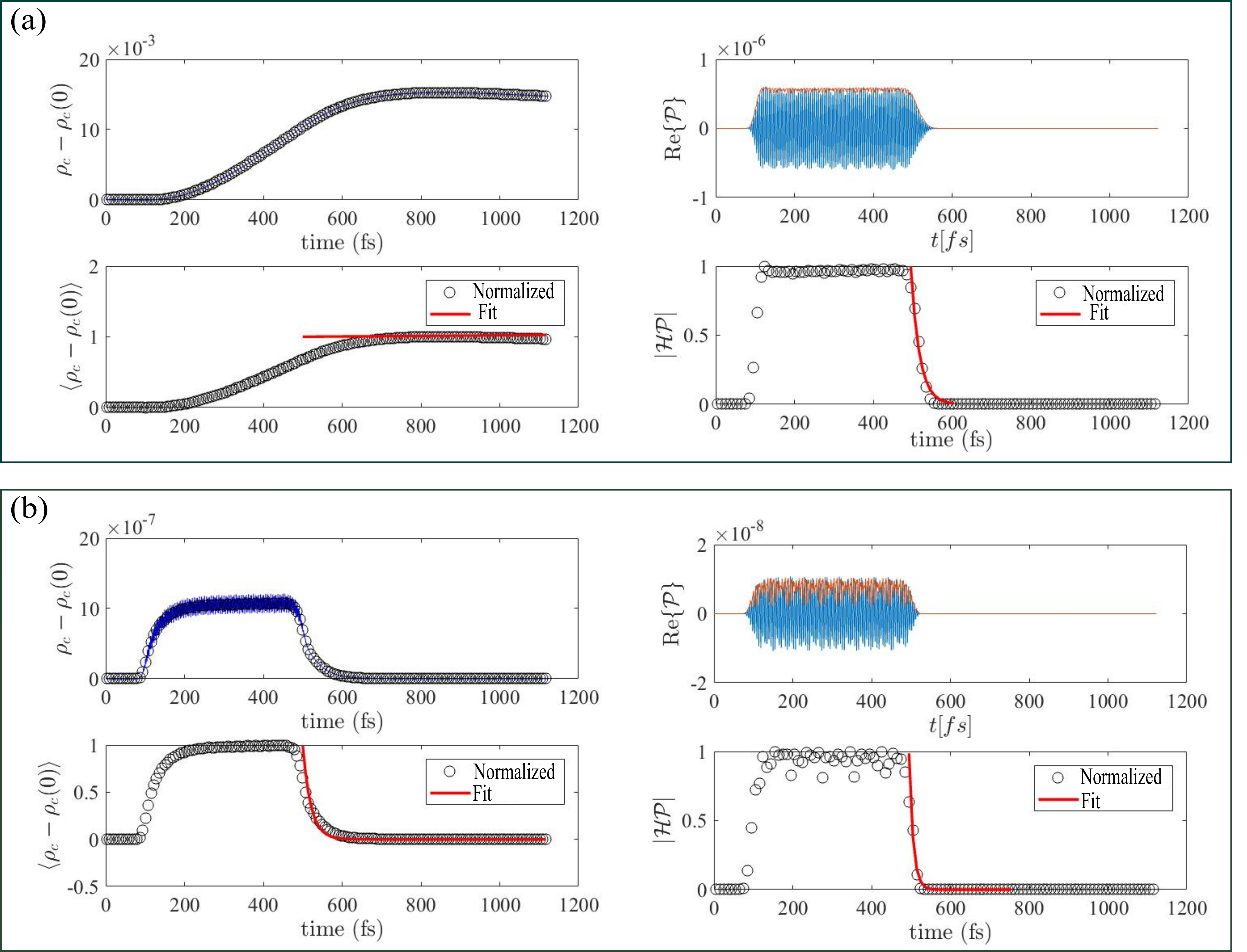}
\caption{\small Extraction of the microscopic relaxation coefficients for the optical excitation of $\hbar\omega=800\mathrm{meV}$ at (a) $\hbar v_F k=50\mathrm{meV}$(low energy) and (b) for $\hbar v_F k=500\mathrm{meV}$ (high energy) points.}
\label{fig:extraction_gamma-800meV}
\end{figure}

\begin{enumerate}[(I)]
\item
Use a Hilbert transform to capture the envelope of the microscopic polarization.
\item
Use a low-pass filter to get rid of higher harmonics in the polarization envelope as well as the population pulsations.
\item
Curve-fit the decaying part of the dynamics to an exponential function. The decay of the microscopic polarization  is $\gamma^{(2)}_{\vp{k}}$ and that of the population is
$\gamma^{(1)}_{\vp{k}}$. 
\end{enumerate}
Figures~\ref{fig:extraction_gamma-80meV} and \ref{fig:extraction_gamma-800meV} display the steps followed to extract the microscopic relaxation coefficients for the pulse excitation of $\hbar\omega=80\mathrm{meV}$ and $800\mathrm{meV}$, respectively. It is assumed that the amplitude of the electric fields for the both optical pulses are $E_0=10^6\mathrm{V/m}$.  The duration of the pulses have been selected to be $800\mathrm{fs}$ and $400\mathrm{fs}$ for the cases of $80 \mathrm{meV}$ and $800\mathrm{meV}$, respectively.

\begin{figure*}[!b]
\centering
\begin{subfigure}[]{
\includegraphics[width=0.4\linewidth]{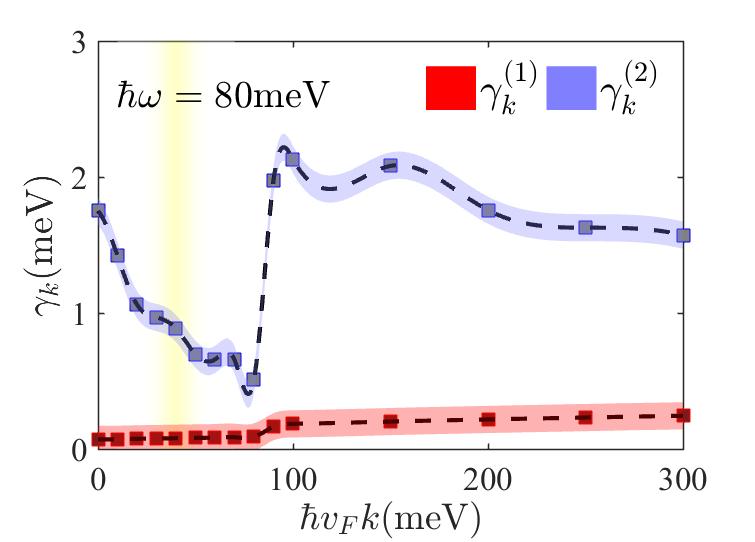} 
\label{fig:gamma_80_app} }
\end{subfigure}
\hfill
\begin{subfigure}[]{
\includegraphics[width=0.4\linewidth]{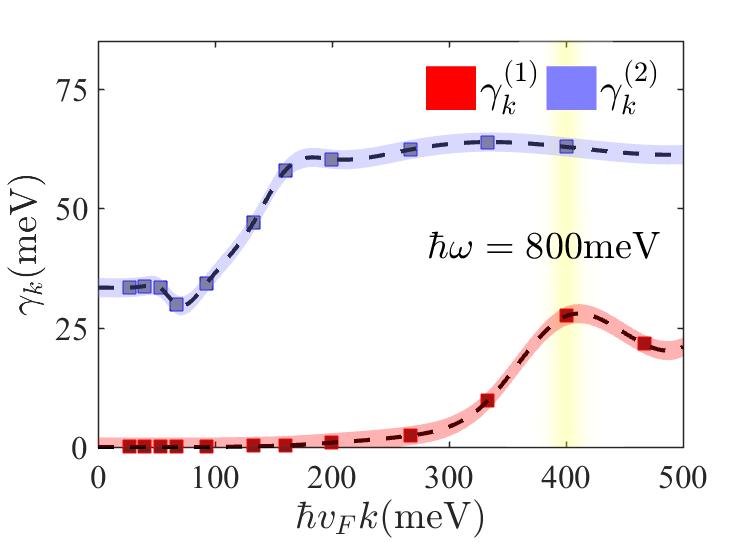} 
\label{fig:gamma_800_app} }%
\end{subfigure}

\caption{
\small k-dependent relaxation coefficients $\gamma^{(1)}_{\vp{k}}$ and $\gamma^{(2)}_{\vp{k}}$ for undoped graphene. The yellow shaded regions are the zero-detuning regions.}
\label{fig:gammak_app}
\end{figure*}

The results of the calculations for a few distinct points over the k-space are shown in Fig.~\ref{fig:gammak_app}. As  expected, the low-energy excitation offers significantly slowed relaxation dynamics. It is also worth noting that, unlike the coherence dephasing coefficient $\gamma^{(2)}_{k}$, the coefficient $\gamma^{(1)}_{k}$ tends to be negligibly small in the vicinity of the Dirac point. The red solid lines are the fitted exponential (decay) functions. For the microscopic polarization, the envelope of the curve needs to be captured where $\mathcal{H}\mathcal{P}$ stands for the Hilbert transform of the polarization. The calculations are shown for two distinct points over the reciprocal space (low energy around the Dirac point and a high energy slightly higher than the zero detuning region)

\newpage
\section{Theoretical Study of Pump-Probe Experiments in Graphene}
The frequency of the pump and probe are $\omega_c$ and $\omega_p$, respectively, and the intensity of the pump laser is denoted by $I_c$.  The intraband absorption profile of the probe field is obtained by plugging the steady-state population difference $\mathcal{N}^{st}_{\vp{k}}(I_c)$ into the linear response theory. The  corresponding intraband linear conductivity tensor in the presence of the pump $I_c$ reads

\begin{figure*}[t]
\includegraphics[width=0.3\textwidth]{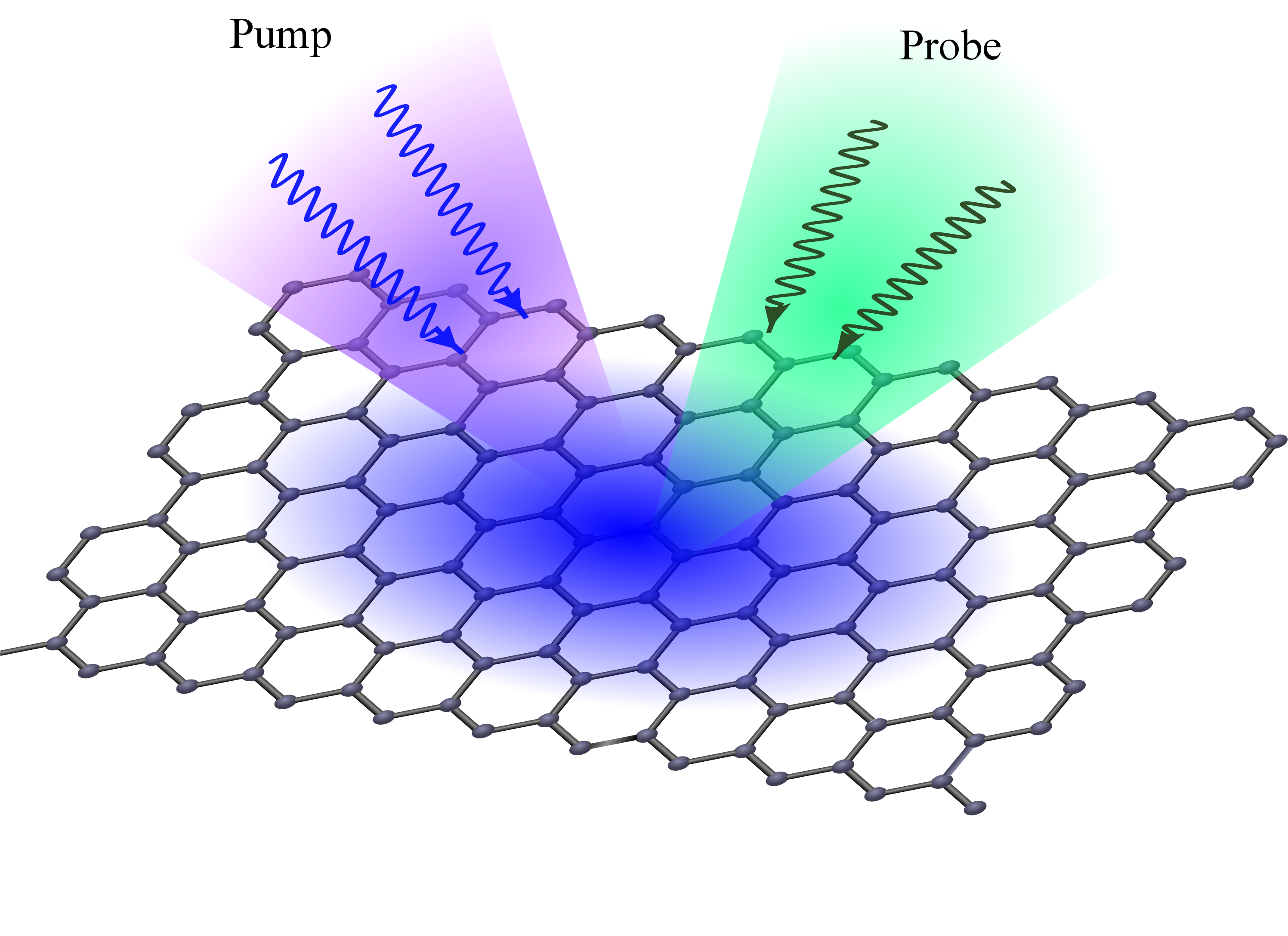}
\caption{Schematic of the two-frequency pump-probe experiment.
}
\label{fig:pump-probe-1}
\end{figure*}

\begin{equation}
\left.\overline{\overline{\Sigma}}_{intra}\right|_{I_c}(\omega_p) \approx
\frac{e^2v_Fg_sg_vD}{i\omega_p+\Gamma}\iint_{\mathrm{RBZ}} \mathrm{d}\vp{k}^{2}\A{\vp{k}}\cdot\nabla_{\vp{k}}\T{\mathcal{N}}^{st}_{\vp{k}}(I_c)
\end{equation}
where $g_s$ and $g_v$ are the spin and valley degeneracy factors, respectively, and $D=1/(2\pi)^2$ is the density of the states in the reciprocal space. 
The integral goes over the reduced reciprocal space where the Dirac dispersion is valid. A more rigorous treatment would require including additional contributions accounting for the nonlinear frequency mixing due to the pure intraband process. However, the above formulation is adequately accurate for large enough pump frequencies. The interband conductivity consists of two contributions, namely incoherent and coherent terms~\cite{Boyd1988}. Similarly to the intraband part, the incoherent contribution simply results  from reduction of the population difference over the reciprocal space. The coherent term, however, enters due to population pulsation at the beat frequency $\omega_c-\omega_p$. It is argued that the coherent term involves interference between the pump and probe fields \cite{Boyd1988}. The population pulsation followed by absorption of a second photon from the pump field acts mathematically as an additional \textit{complex-valued} contribution to the population difference denoted by  $\delta\mathcal{N}^{puls}$. The overall pump induced interband conductivity tensor is 
\begin{align}
\left.\overline{\overline{\Sigma}}_{inter}\right|_{I_c}(\omega_p) &=-\frac{e^2}{\hbar}v_Fg_sg_vD
 \iint_{\mathrm{RBZ}} \mathrm{d}\vp{k}^{2}\left\{
\A{\varphi}_{\vp{k}}\A{\varphi}_{\vp{k}} \frac{1}{k}\mathcal{L}_{\vp{k}}(\omega_p)\left[\T{\mathcal{N}}_{\vp{k}}^{st}(I_c)+\delta\mathcal{N}^{puls}_{\vp{k}}(I_c)\right]
\right\}\nonumber\\
&=\overline{\overline{\Sigma}}_{incoh}+\overline{\overline{\Sigma}}_{coh}
\end{align}
The complex-valued Lorentzian $\mathcal{L}_{\vp{k}}(\omega)$  is abbreviated as
\begin{equation}
\mathcal{L}_{\vp{k}}(\omega)\triangleq\frac{1}{\gamma^{(2)}_{\vp{k}}+i\Delta_{\vp{k}}}
\end{equation}
Here $\Delta_{\vp{k}}=\omega-\varpi_\vp{k}$ denotes the detuning of the two-level system at $\vp{k}$ with respect to the excitation.
The complex population $\delta\mathcal{N}^{puls}_{\vp{k}}$ is
\begin{equation}
\delta\mathcal{N}^{puls}_{\vp{k}}(I_c)=-\T{\mathcal{N}}_{\vp{k}}^{st}(I_c)
\frac{\left[\mathcal{L}_{\vp{k}}(\omega_p)+\mathcal{L}^*_{\vp{k}}(\omega_c)\right]\abs{\T{\Phi}^c_{\vp{k}}}^2}{2\gamma^{(1)}_\vp{k}-2i(\omega_c-\omega_p)+\left[\mathcal{L}_{\vp{k}}(\omega_p)+\mathcal{L}^*_{\vp{k}}(2\omega_c-\omega_p)\right]\abs{\T{\Phi}^c_{\vp{k}}}^2}
\end{equation}
The validity of the theory requires that the coherence length of the pump laser be  much larger than the wavelength of the probe field.

\begin{figure*}[!h]
\centering
\begin{subfigure}[]{
\includegraphics[width=0.45\linewidth]{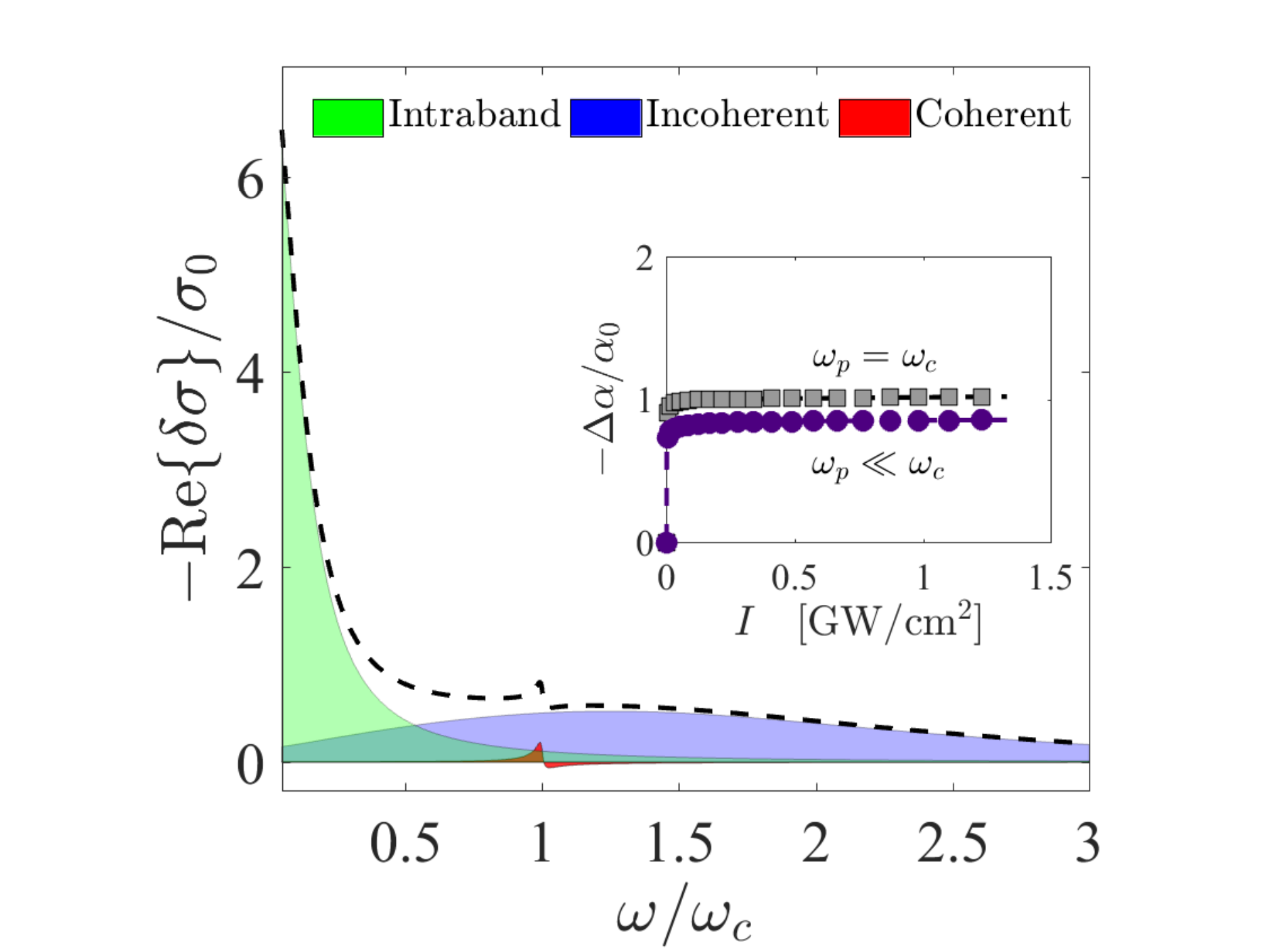} 
\label{fig:pump-probe80} }
\end{subfigure}
\hfill
\begin{subfigure}[]{
\includegraphics[width=0.45\linewidth]{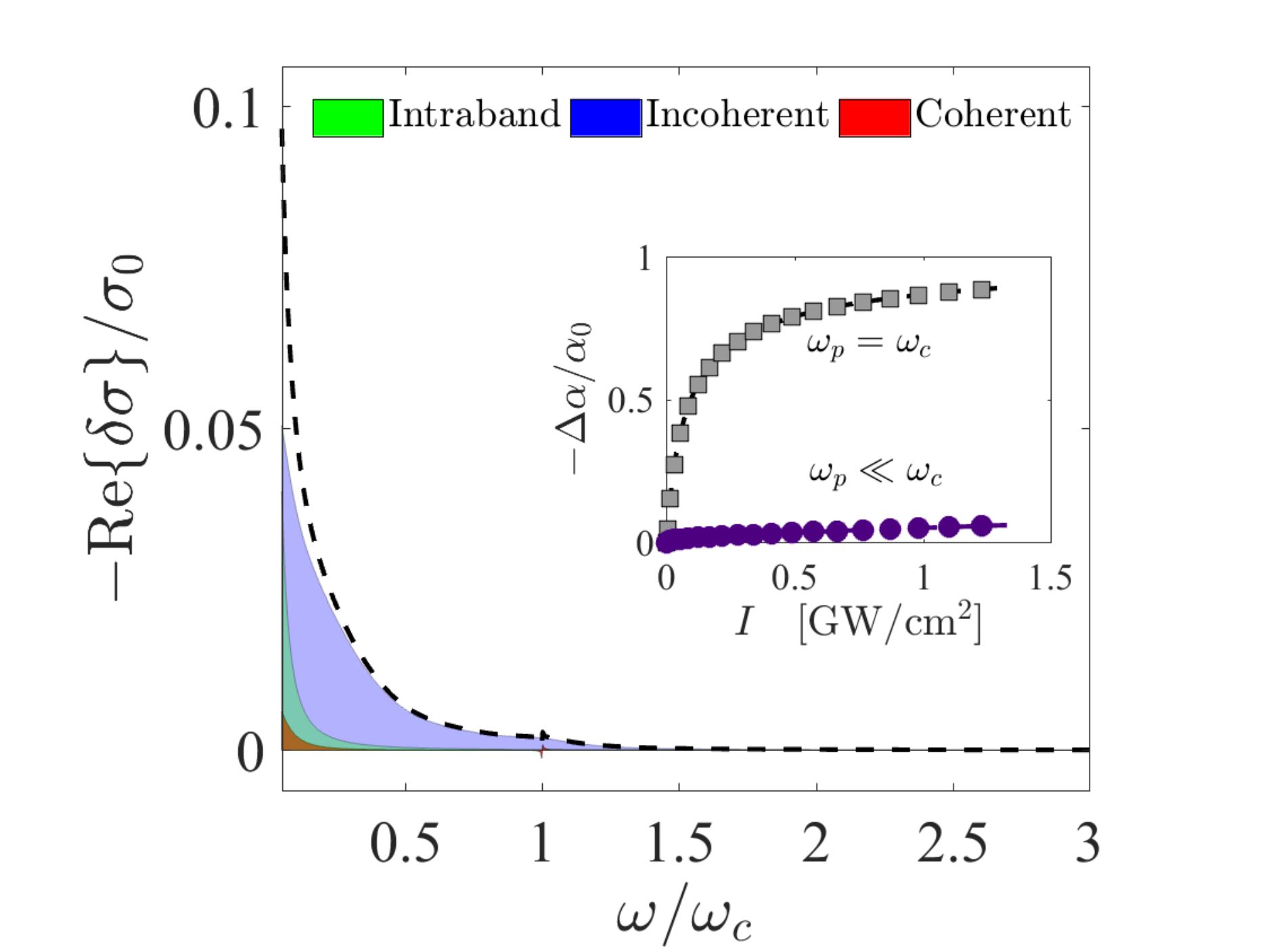} 
\label{fig:pump-probe800} }%
\end{subfigure}

\caption{\small change in the diagonal element of the conductivity  at the probe frequency $\omega_p$ ($\delta\sigma (\omega_p)$) due to the presence of the pump field at the frequency of (a) $\hbar\omega_c=80\mathrm{meV}$ and (b) $\hbar\omega_c=800\mathrm{meV}$. The conductivity is normalized to $\sigma_0=e^2/4\hbar$. 
Variations of the absorption coefficient of graphene (normalized to its intrinsic value) versus the pump intensity for  two frequencies  of the probe field are shown in the insets of the figures. 
}
\label{fig:pump-probe}
\end{figure*}

\section{Nonperturbative Kerr-Type Nonlinearity of Graphene\label{app:Nonperturbative}}

A nonperturbative formulation of the Kerr effect in graphene is obtained by taking the steady-state population difference as the optically modified inversion. In the steady-state analysis, essentially only one-photon processes are retained. The impact of higher-order effects including two-photon absorption (TPA) is then incorporated into the response function via additional complex contributions to the population difference. By substituting \textit{the effective complex population difference} into the linear response theory, one obtains the  induced interband nonlinear current as

\begin{equation}
\vec{NL}_1(\omega)=-\frac{e^2}{\hbar}v_Fg_sg_vD
 \iint_{\mathrm{RBZ}} \mathrm{d}\vp{k}^{2}\left\{\A{\varphi}_{\vp{k}} (\A{\varphi}_{\vp{k}}\cdot\vp{E}) 
\frac{1}{k}\mathcal{L}_{\vp{k}}(\omega)\left[\T{\mathcal{N}}_{\vp{k}}^{st}+\delta\mathcal{N}^{TPA}_{\vp{k}}+\delta\mathcal{N}^{B}_{\vp{k}}-\mathcal{N}_{\vp{k}}^{eq}\right]
\right\}
\end{equation}
where the complex function $\delta\mathcal{N}^{TPA}_{\vp{k}}$ accounts for the population oscillations at the second harmonic of the excitation. In the dressed-state picture, TPA is conceived as the absorption of two subsequent photons via a virtual state. In the density-matrix formulation, that contribution results in the oscillation of the population difference at the second harmonic. By expanding the inversion as $\mathcal{N}_{\vp{k}}\approx \T{\mathcal{N}}^{st}_{\vp{k}}+(\mathcal{N}^{(2)}_{\vp{k}}\exp(i2\omega t)+c.c.)$ and plugging this term into the SBEs, one obtained the complex contribution to the inversion as
\begin{equation}
\delta\mathcal{N}^{TPA}_{\vp{k}}=-\frac{1}{2}\mathcal{L}_{\vp{k}}(\omega)\frac{|\T{\Phi}_{\vp{k}}|^2}{\gamma^{(1)}_\vp{k}+2i\omega}\T{\mathcal{N}}^{st}_{\vp{k}}
\end{equation}
The second contribution $\delta\mathcal{N}^{B}_{\vp{k}}$ enters due to intraband dynamics or Boltzman type transport. According to Eq.~\eqref{eq:Boltzman}, the intraband dynamics can be incorporated by displacing the density matrix in k-space by $\Delta\vp{k}$. As a result of two subsequent translations conducted by two conjugate  fields with the frequency of $\omega$ and $-\omega$, the steady-state population difference experiences a minor change denoted by $\delta\mathcal{N}^{B}_{\vp{k}}$. By Taylor expanding  the population, one obtains

%

\bibliographystyle{apsrev4-1}

\end{document}